\documentclass[12pt]{article}

\usepackage[round]{natbib}

\usepackage{url}
\usepackage{graphics}
\usepackage{multirow}
\usepackage{rotating}
\usepackage{subfiles}
\usepackage{tikz}
\usepackage{chapterbib}
\usepackage{lscape}
\usepackage{pdflscape} 
\usepackage{afterpage}
\usepackage[ansinew]{inputenc}
\usepackage[T1]{fontenc}
\usepackage{lmodern}
\usepackage[english]{babel} 
\usepackage[babel]{csquotes}
\usepackage[bottom]{footmisc} 
\usepackage{booktabs,tabularx,tabu}
\usepackage{array,longtable,multirow}
	\newcolumntype{Y}{>{\centering\arraybackslash}X}
\usepackage{graphicx}
\usepackage[font=small,labelfont=bf,singlelinecheck=false]{caption} 
\usepackage{amsmath}
\usepackage{float}
\usepackage{xcolor}
\usepackage[doublespacing]{setspace} 

\usepackage[breaklinks, linktocpage, bookmarksnumbered=true, colorlinks=true, linkcolor=cyan, citecolor=cyan, urlcolor=cyan, pdftitle={Ph.D. Thesis},  pdfauthor={Alexander Jordan}]{hyperref}

\usepackage{geometry}
\usepackage{multicol}		
\usepackage{fancyhdr}
\definecolor{chameleongreen1}{RGB}{98,189,25}
\renewcommand*{\thefootnote}{\fnsymbol{footnote}}

\fancypagestyle{lscapedplain}{%
  \fancyhf{}	
  
	\fancyfoot{%
    \tikz[overlay]
      \node[outer sep=2cm,above,rotate=90] at (current page.east) {\thepage};}
}

\usepackage{titlesec}

\large

\makeatletter
\def\@footnotecolor{red}
\define@key{Hyp}{footnotecolor}{%
 \HyColor@HyperrefColor{#1}\@footnotecolor%
}
\def\@footnotemark{%
    \leavevmode
    \ifhmode\edef\@x@sf{\the\spacefactor}\nobreak\fi
    \stepcounter{Hfootnote}%
    \global\let\Hy@saved@currentHref\@currentHref
    \hyper@makecurrent{Hfootnote}%
    \global\let\Hy@footnote@currentHref\@currentHref
    \global\let\@currentHref\Hy@saved@currentHref
    \hyper@linkstart{footnote}{\Hy@footnote@currentHref}%
    \@makefnmark
    \hyper@linkend
    \ifhmode\spacefactor\@x@sf\fi
    \relax
  }%
\makeatother
\hypersetup{footnotecolor=black}

\begin{document}

%
%

\thispagestyle{empty}
	\begin{center}
		{\Large The pain of a new idea:}\\ 
		\vspace*{0.5cm}
		{\large Do Late Bloomers response to Extension Service in Rural Ethiopia}\\
		\vspace*{0.5cm}
		{Alexander Jordan$^{a}$ Marco Guerzoni $^{a,b,}$}\,\footnote{\,Corresponding author at: Department of Economics and Statistics  ``Cognetti de Martiis'', Lungo Dora Siena 100A, 10153 Torino, Italy. Tel.: +39 011 670 9383. E-mail address: {\url{alexander.jordan@unito.it}}}\\
		\vspace*{0.5cm}
		{$^{a}$ \scriptsize Department of Economics and Statistics ``Cognetti de Martiis'', University of Turin, Italy} \\
		{$^{b}$ \scriptsize ICRIOS, Bocconi University} \\
		\end{center}
%

\begin{onehalfspacing}
	\begin{abstract}
	\noindent 
	The paper analyses the efficiency of extension programs in the adoption of chemical fertilisers in Ethiopia between 1994 and 2004. Fertiliser adoption provides a suitable strategy to ensure and stabilize food production in remote vulnerable areas. Extension services programs have a long history in supporting the application of fertiliser. However, their efficiency is questioned. In our analysis, we focus on seven villages with a considerable time lag in fertiliser diffusion. Using matching techniques avoids sample selection bias in the comparison of treated (households received extension service) and controlled households. Additionally to common factors, measures of culture, proxied by ethnicity and religion, aim to control for potential tensions between extension agents and peasants that hamper the efficiency of the program. We find a considerable impact of extension service on the first fertiliser adoption. The impact is consistent for five of seven villages. 

	\vspace*{0.3cm}
	\end{abstract}
\end{onehalfspacing}
\pagenumbering{arabic}

\setcounter{footnote}{0}
\renewcommand*{\thefootnote}{\arabic{footnote}}

\section{Introduction}

The eradication of extreme poverty and hunger plays an essential role for the development of Sub-Saharan Africa \citep{UnitedNations2014}. The Ethiopian strategy, to promote proper nutrition and stable food supply, builds on large investments in the agricultural sector in order to spur labour productivity \citep{Rashid2013a}. Current figures indicate the success of that policy \citep{Dorosh2013}. Despite the fact that Ethiopia was placed under the \textit{alarming hunger} category by the Global Hunger Index\footnote{The Global Hunger Index combines undernourishment, child underweight and child mortality as equally weighted indicators \citep{IFPRI2014}.}, it was able as a nation to meet the MDG's hunger target \citep{FAO2014} and to cut the relative and absolute size of its starving population \citep{UN-OHRLLS2014}. Since the beginning of the 1990s, annual fertiliser consumption grew from 140,000 tons up to 650,000 tons in 2012 \citep{Rashid2013a} and fertiliser usage increased by 180 percent between 1993 and 2005 \citep{UNDP2010}. The main channel to promote fertiliser usage in rural areas are extension service programmes \citep{Byerlee2007}. These programmes aim to support the application of modern inputs like fertiliser or improved seeds and also teach new agricultural practices or serve as a source of credit for agricultural equipment. However, their efficiency is questioned by the literature as application rates remain low \citep{Byerlee2007,Kassie2009,Spielman2011a}, resulting, in large stocks due to an excess supply of fertiliser \citep{Abrar2004,Rashid2013a}.
	
Therefore, the paper aims to contribute to the discussion about the effectiveness of extension service programmes. Recent literature by \citet{Krishnan2014} and \citet{Nisrane2011} found large positive impacts of extension service on fertiliser adoption and agricultural output between 1994 and 2004 but rarely any impact after 2004. We confirm the results of \citet{Krishnan2014} and observe a notable positive impact of extension service on fertiliser adoption between 1994 and 2004\footnote{In contrast to \citet{Krishnan2014}, we lack the necessary information regarding extension service participation between 2004 and 2009.}. Unlike previous works on the topic, we use the pure definition of adoption and focus only on households without prior knowledge of fertiliser usage. The underlying justification lies in the recurrent necessity of fertiliser usage. Since the decision to use fertiliser occurs repeatedly, previous personal experience of households influences the repeated decision after the initial adoption. Hence, the observed diminishing impact of extension service as source of information on fertiliser usage after 2004 may be attributed to its redundancy as the vast majority of households has already applied fertiliser at least once and does not require further extensions service but, instead, relies on close neighbours for knowledge exchange and for a discussion regarding continued usage, as shown by \citet{Krishnan2014}. Our analysis identifies the effect of extension service on first adopters and provides evidence for the importance of the programme in areas where fertiliser had not previously been used. Thus, we measure the pure effect of extension service and investigate whether extension agents fulfil their intended function of raising awareness about the technology and spurring diffusion. In addition, the study highlights potential obstacles to the performance of development agents from a cultural vantage point. \par

The remainder of the paper is organized as follows: \hyperref[Literature]{Section\,\ref{Literature}} briefly reviews the history of extension programmes in Ethiopia and the discussion about the efficiency of such programmes. \hyperref[DataMatch]{Section\,\ref{DataMatch}} presents the data selected from the Ethiopia Rural Household Survey (ERHS) and investigate determinants of late bloomers. \hyperref[Method]{Section\,\ref{Method}} describes the matching technique applied to analyse the impact of extension services on fertiliser adoption. \hyperref[Result]{Section\,\ref{Result}} presents the results and discusses the robustness and heterogeneity of the effect. Finally, \hyperref[ConclusionMatch]{Section\,\ref{ConclusionMatch}} closes the paper with a summary and an outlook for further research.
\par	

\section{Extension service in Ethiopia}
\label{Literature}

Soil degeneration, erosion and deforestation are common problems in Ethiopia and complicate a sufficient food supply for the growing population. In addition, current farming practices and technologies are outdated and agricultural productivity lags far behind \citep{Diao2007, AGRA2014}. A potential solution to foster agricultural productivity is seen in the promotion of extension service programmes. These development programmes aim to provide information and awareness of modern technologies and suitable practices. They aim to close knowledge gaps between farmers and research in order to increase yields and exploit the agricultural potential to ensure sustainable food production. Positive examples of functioning extension services can be found worldwide, e.g., Pakistan \citep{Ali2013}, Uruguay \citep{Maffioli2013} or Nigeria \citep{Fabiyi2015} amongst others. \par

In Ethiopia, agricultural extension service programmes first occurred in the 1950th \citep{Belay2003}. Since then, programmes with different design and scope like the Minimum Package Project I\,\&\,II (from 1971 to 1974 and 1981 to 1985) or the Peasant Agriculture Development Extension Programme (PADEP, launched in 1985) have been in place \citep{Belay2003,Belay2004}. These programmes have been generally inefficient due to their top-down approach and the non participatory nature \citep{Belay2003, Mossie2015}. More recent approaches like Sakawa Global 2000 (SG2000, 1993) or the participatory demonstration and training extension system (PADETES, 1995) aimed to resolve these shortcomings. However, evaluation results are mixed. While \citet{Bonger2004} or \citet{Spielman2011a} find hardly any positive impacts for income and production, \citet{Dercon2009} attributes a decrease of 9.8 percentage points in poverty to the work of the extension agents. With respect to the use of new technologies, \citet{Kassie2009} discovers a positive influence of extension agents on the adoption decision. Also \citet{Nisrane2011}, \citet{Ragasa2013} and \citet{Krishnan2014} reveal positive and large effects on fertiliser usage for the time since the launch of the PADETES programme until 2004 but barely any impact afterwards \citep{Nisrane2011,Krishnan2014}. \par 

Common problems relating to fertiliser diffusion are seen in the fluctuation of prices as well as uncertainties in production and consumption \citep{Kassie2009,Dercon2011}. However, \citet{Abrar2004} show that prices are not the major obstacle to fertiliser diffusion. Instead, they suspect that it is the lack of the access to fertiliser and credits that hinder diffusion \citep{Abrar2004}. The potential shortcomings in access to fertiliser and credit have been counterbalanced by an massive increase in the quantity of extension workers, who are responsible for the provision of credit and fertiliser, besides agricultural training and other governmental tasks \citep{Davis2010}.\par

In general, the main problem is not seen in the efficiency of extension programmes but in the low and gender biased participation \citep{Davis2010,Ragasa2013}. Low participation levels may also explain the higher importance of neighbours for fertiliser adoption observed by \citet{Krishnan2014} as the propensity to adopt does not depend solely on the contact with the agent but on the perceived usefulness of their advices \citep{Ragasa2013}. Even though, low quality recommendations and missing consideration of indigenous knowledge have been cited as weak points of extension service \citep{Belay2003,Davis2010}, farmers perception may be biased due to the cultural, i.e. ethnic or religious, background of the responsible advisor. \citet{Abebe2016} provide empirical evidence for the works by \citet{DeWeerdt2002} or \citet{Munshi2014} and show farmers preference to collaborate with ethnic or religious adherents. On the one hand, norms and values that are shared by cultural allies, allow the establishment of trust and provide a common sense which can facilitate cooperation \citep{Platteau1994,Fafchamps2006,Karlan2009}. On the other hand, common norms in rural societies can oblige individuals to act against their own interests \citep{Fehr1997,Platteau2009}, like the cultural constraint of using microcredits in certain parts of Ethiopia \citep{Davis2010}. Hence, the participation in and the perceived usefulness of the extension programme may be subject to a cultural bias if the ethnic or religious background of the agent does not fit the prevailing norms of the rural society. A cultural mismatch is not unlikely as Ethiopia is home to more than 80 ethnic groups \citep{Census2007}, and intermixing across ethnic territories occurs frequently \citep{Bekele2003}. Besides conflicting ethnic norms, language barriers may appear as each ethnic group inherits its own tongue, leading to language distinctions even between areas that are just a short distance apart, especially in the multi-ethnic Southern Nations, Nationalities, and Peoples Region. Thus, linguistic obstacles can impede the development of trust due to communication gap and simultaneously lacking information and knowledge exchange \citep{Breuer2012}. \par

The presented literature does not distinguish between the first adoption and repeated adoptions. Hence, the performance of extension agents is measured independently from farmers' prior knowledge and experiences about fertiliser usage. Yet, prior experiences influence personal attitudes towards available technology and the perceived value of extension agent recommendations. Even the best advice may fail if unpredictable environmental circumstances prevent fertiliser to augment yields and impairs the reputation of the development agent. Thus, the mixed results and vanishing importance of extension service for fertiliser adoption may be partially attributed to the impure adoption definition besides previously mentioned factors. Therefore, the following sections will investigate the performance of extension service programmes among inexperienced farmers.

\section{Data}
\label{DataMatch}
\subsection{Data selection}

The paper uses a sub-sample of seven Peasant Association (PA)\footnote{Namely the Peasant Associations Adado, Dinki, Doma, Geblen, Haresaw, Imdibir and Shumsheha.} and 643 households from the Ethiopia Rural Household Survey (ERHS) which originally covers 1477 rural households in 15 Peasant Association between 1994 and 2009\footnote{The questionnaire of the 2009 round of the survey lacks the specific questions about extension service participation between 2004 and 2009 and the information is not identifiable in the raw data. Hence, the last survey round is excluded from the analysis.}. The selection reflects the necessity to consider prevalent diffusion levels\footnote{Hereby, the diffusion of fertiliser presents the share of households within a society, that have applied fertilisers at least once. The author is aware that the initial adoption of fertiliser does not imply a persistent usage and that the decision to apply fertiliser occurs repeatedly over time which results in a fluctuation of fertiliser application rates between years.} in order to properly analyse the effect of extension services. As seen in the first column of \hyperref[Adopttime]{Table\,\ref{Adopttime}} fertiliser dissemination among all PAs reveals substantial differences in diffusion levels for the survey baseline year 1994. While the villages in the lower section of \hyperref[Adopttime]{Table\,\ref{Adopttime}} barely show a proper initialization of the diffusion process, fertiliser was used at least once by more than 60\% of households in villages from the upper section of \hyperref[Adopttime]{Table\,\ref{Adopttime}}. In addition, the villages have disparities with respect to their chronological starting points as seen in column 2. \par

\begin{table}[H]
\begin{footnotesize}
\caption[Fertiliser Diffusion across Peasant Associations]{\centering{\textit{Fertiliser Diffusion across Peasant Associations}}} 
\label{Adopttime}
\begin{tabularx}{\textwidth}{@{} ll*{4}{Y} @{}}
\toprule                                             
& Pesant Association				& Diffusion Level in 1994  & First Adoption (Year) & Diffusion Level by 2009 
\\ 
\midrule

\parbox[t]{1mm}{\multirow{8}{*}{\rotatebox[origin=c]{90}{\textit{Mature Adopter}}}} 
& Aze Deboa 				& 98.67 &	1964 			& 100 		\\ 
& Trirufe Ketchema 	& 94.12 & 1958			& 99.01 \\ 
& Gara Godo 				& 91.97 &	1965			& 98.95 \\ 
& Sirba na Goditi 	& 89.69 & 1958			& 92.78 	\\ 
& Debre Berhan 			& 89.13 &	1964			& 97.28 \\ 
& Yetmen 						& 86.88 &	1977			& 91.80 \\ 
& Koro Degaga 			& 70.64 &	1977			& 96.33 \\ 
& Adele Keke 				& 61.85 & 1965			& 89.69 \\ 
\midrule 
\parbox[t]{1mm}{\multirow{7}{*}{\rotatebox[origin=c]{90}{\textit{Late Bloomers}}}}
& Dinki 						& 12.64 &	1984			& 60.91 \\ 
& Haresaw 					& 9.52 & 1989			& 80.95 	\\ 
& Doma 							& 5.4 & 1994			& 67.56 	\\ 
& Geblen 						& 4.54 &	1992			&	40.90 	\\ 
& Shumsheha 				& 0.67 & 1994			&	28.37 	\\ 
& Imdibir 					& 0 & 1995			& 35.82 	\\ 
& Adado 						& 0 & 1996			& 14.61 	\\ 

\bottomrule 	\\	\vspace{-5ex} 				
\end{tabularx} 
\textit{Source:} Author's calculations based on ERHS.\\  
\textit{Note:} The diffusion levels express the percentual share of fertiliser adopters per village. Columns 2\,\&\,4 include adoptions occurred in 1994\,\&\,2009. Column 3 presents the year of first fertiliser adoption in each PA. \\
\end{footnotesize}
\end{table}

These initial differences in fertiliser diffusion suggest to not analyse the performance of extension service among all PAs. On the one hand, precise information about extension service participation before 1994 is missing, makes it impossible to link the first adoption of fertiliser to the presence of extension service in the years prior the launch of the survey. On the other hand, fertiliser usage after 1994 in villages with high diffusion levels may not be solely attributed to the extension service but also to the prevalent knowledge and experience about fertiliser usage in the local environment like shown by \citet{Krishnan2014}. Thus, the analysis will focus on households without access to prior experience about fertiliser in the local environment\footnote{Even though there where few cases of fertiliser being adopted in Dinki, Haresaw and Geblen before 1994, they are marginal in numbers and should have a negligible impact on the shareable mindset of these PAs.}. \par

\begin{figure}[H]
\begin{footnotesize}
\caption[Year of First Adoption by PA and Diffusion of fertiliser]{\centering{\textit{Year of First Adoption by PA and Diffusion of fertiliser}}} 
\label{Adopter}
\includegraphics[scale=0.54]{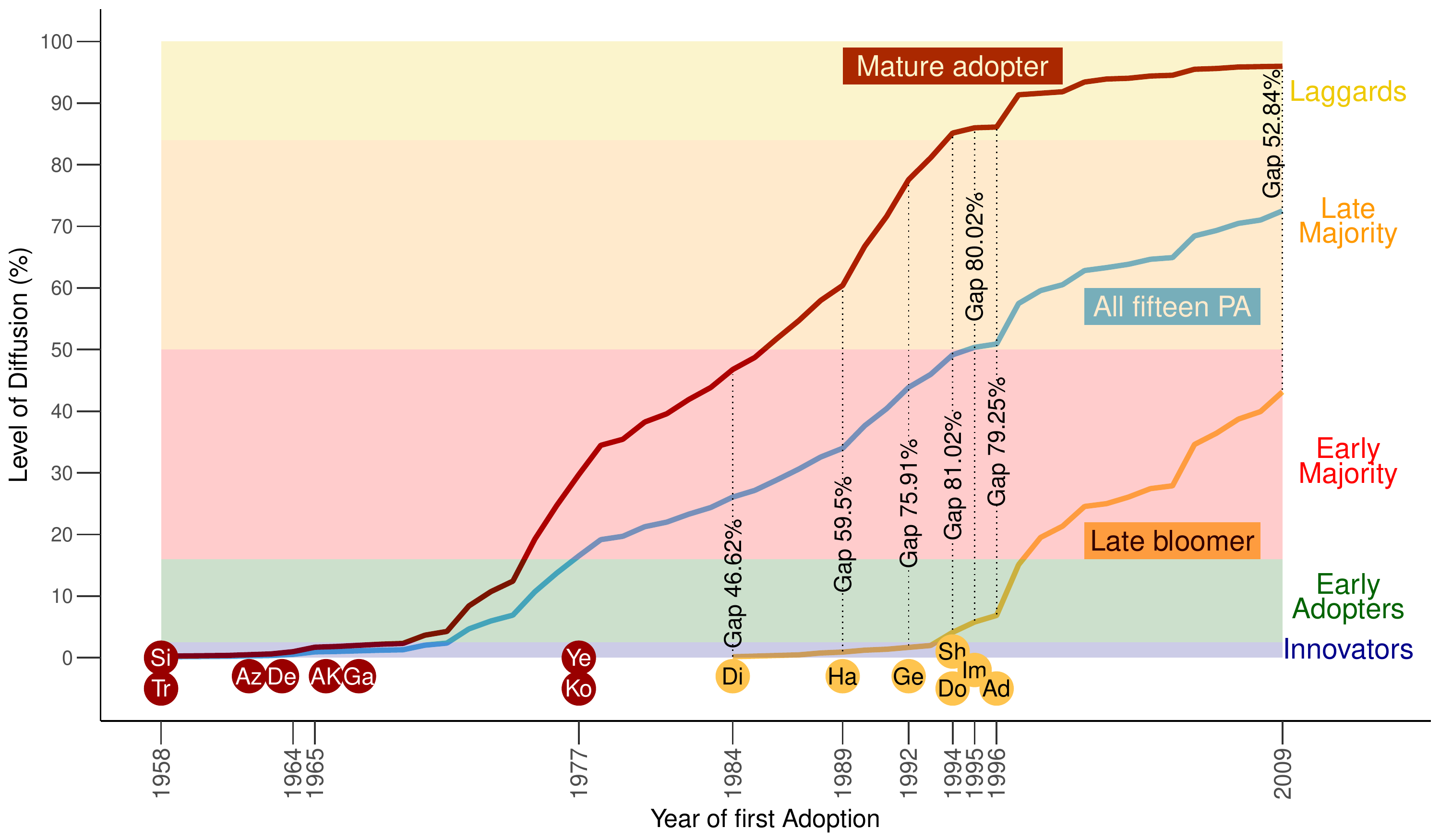}
\textit{Source:} Author's calculations based on ERHS and applied on the adopter categorization by \citet{Rogers2003}.\\  
\textit{Note:} The x-axis depicts years when fertiliser usage started in each PA (See also \hyperref[Adopttime]{Table\,\ref{Adopttime}}). The blue curve illustrates the total fertiliser diffusion of all PA, the dark red curve considers only PAs that start adoption prior 1984 and the dark yellow curve depicts diffusion for PAs with adoption initialization in 1984.\\
\end{footnotesize}
\end{figure}

However, understanding the vast differences in diffusion levels in 1994 and the time gaps of fertiliser start off is useful as these outcomes may be based on systematic dissimilarity of key variables. Hence, the PAs are split into two groups, namely mature adopters and late bloomers, to track down the causes. The group labels are inspired by \citet{Rogers2003} adopter categories\footnote{\citet{Rogers2003} classifies the categories based on the standard deviations from the average time of adoption. As we deal with censoring and not complete adoption, our categorization relies on diffusion levels.} and \hyperref[Adopter]{Figure\,\ref{Adopter}} illustrates the reason for complying with his classification. Notably, a huge chronological gap appears between the first overall adoption in 1958 and the take up in fertiliser usage among the late bloomers. A comparison between the blue and dark red lines reveals the hampering effect of delayed fertiliser diffusion by the late bloomer group on the overall fertiliser diffusion. The late bloomers face their first adoption not until the diffusion among mature adopters already reached more than 46\% and the vast majority of adoptions takes even place when fertiliser diffusion among the mature adopters already reaches the theoretical tipping point of the diffusion curve. The corresponding adopter categories show that households that appear locally as \textit{Innovators} or \textit{Early Adopters} among the late bloomers would mainly represent the \textit{Late Majority} or \textit{Laggards} in the mature adopter group (except for the first adopters in Dinki).  \par 

Interestingly, the time gap of 26 years between initial adoption in both groups checked against the time gap resulting from the final diffusion level of late bloomers (47.02\%) matched at its equivalent on the mature adopter diffusion curve (in 1984/1985), reveals only a marginal decrease of 1.5 years (26 years vs. 24.5 years). In other words, late bloomers did not substantially catch up or benefit from their late fertiliser uptake, suggesting missing spillovers from national fertiliser diffusion to local dissemination. \par 

The comparison of selected household specific variables, known to be influential factors for agricultural technology diffusion by \citet{Feder1985}, between mature adopters and late bloomers in \hyperref[suspects]{Table\,\ref{suspects}} reveal an expected pattern.  \par

\vspace{2ex}
\begin{table}[H]
\begin{footnotesize}
\caption[Comparison of adoption determinants by Adopter Categories]{\centering{\textit{Comparison of adoption determinants by Adopter Categories}}}
\label{suspects}
\begin{tabularx}{\textwidth}{@{} l*{10}{Y} @{}}
\toprule
												&  Mature Adopter 	&  Late Bloomer 	& \multicolumn{7}{c}{Late Bloomer}  \\  
																															\cmidrule(lr){4-10} 		
												&  								& 								&  Adado 	& Dinki	& Doma	& Geblen	& Haresaw	& Imdibir	& Shumsheha	\\
\midrule 
 Plot Size\,(ha)				& 1.53						& 0.76 						& 0.39 		& 1.17	& 1.14 	& 0.22		&	0.47 		&	0.11		&  1.47 \\
\midrule
 Soil Quality						& 1.7 						& 1.85 						& 1.59 		& 1.87	& 1.11	& 2.83 		& 2.44 		& 1.78 		&  1.87 \\
\midrule
 Crop Variety						& 2.57 						& 2.01 						& 1.55 		& 1.79 	& 1.57	& 1.12 		& 1.3 		& 2.98 		&  3.09  \\
\midrule
 Distance\,(km)					& 8.12 						& 10.29 					& 7 			& 11 		& 3 		&	18 			&	16			&	5				&  12   \\								

\bottomrule \\
						
\end{tabularx}
\textit{Source:} Author's calculations based on ERHS \\
\textit{Note:} The comparison presents mean values of adoption determinants in the baseline year 1994. Soil Quality ranges from to one to three with value 1 as indicator for good soil and value 3 for poor soil. Crop variety refers to the average number of different crops on household plots and the distance is measured between village and market. \\
\end{footnotesize}
\end{table}
\vspace{-2ex}

Mature adopters have on average larger plots with better soil quality while farming a larger variety of crops and facing shorter distance to markets\footnote{Since values in \hyperref[suspects]{Table\,\ref{suspects}} correspond to 1994 as reference year, all indicators, apart from distance to market, may be erroneous due to fertiliser uptake before 1994.}. Geblen in particular suffers from the worst local conditions with the largest distance to market, the worst soil quality and the penultimate farm sizes. Nevertheless, Geblen begins fertiliser adoption earlier than relatively more advantageous villages like Adado, Doma, Imdibir or Shumsheha.  \par

At first glance, understanding why certain villages of the late bloomer group reveal chronological differences in initiating the fertiliser diffusion process cannot be exclusively justified by comparing the feasibility of some adoption indicators. Considering the political and environmental circumstances in Ethiopia, two additional explanations should be highlighted. After the fall of the monarchy in 1974 and the establishment of a military dictatorship, the Derg treated northern Ethiopian areas as enemies. According to the supplementary village studies of the ERHS, Geblen and Haresaw were facing a civil war under the Derg regime (1974-1991) that: ``[...] heavily devastated the area economically and ecologically, and socially compounded the natural calamities.'' \citep{Haresaw}. Beside the social insecurity, the notorious famine of 1984 caused the resettlement of households in many areas and the PA of Doma just emerged due to the migration of households from the drought-affected Gamo highlands in 1985 \citep{Doma}. Thus, households of these three PAs were prevented from using fertiliser due to civil war or the harsh living conditions that caused resettlement\footnote{In order to avoid bias from migration, we correct for migration history and do not consider fertiliser adoption prior to land ownership.}. In turn, the historical political and environmental conditions fail to fully explain the low diffusion levels in Dinki and Shumsheha or the total absence of fertiliser usage in Adado and Imdibir in 1994\footnote{Droughts and other weather shocks were severe in any of the 15 PA's of the ERHS and all suffered to various degrees from the Derg regime. However, fertiliser application took place in most locations and only Doma, Geblen and Haresaw seem to be thwarted by the political situation in this particular context.}.
\par

\subsection{Extension service among late bloomers}

\begin{figure}[H]
\begin{footnotesize}
\caption[Contact with extension service and fertiliser adoption conditional on contact]{\centering{\textit{Contact with extension service and fertiliser adoption conditional on contact}}}
\label{Participation}
\includegraphics[scale=0.26]{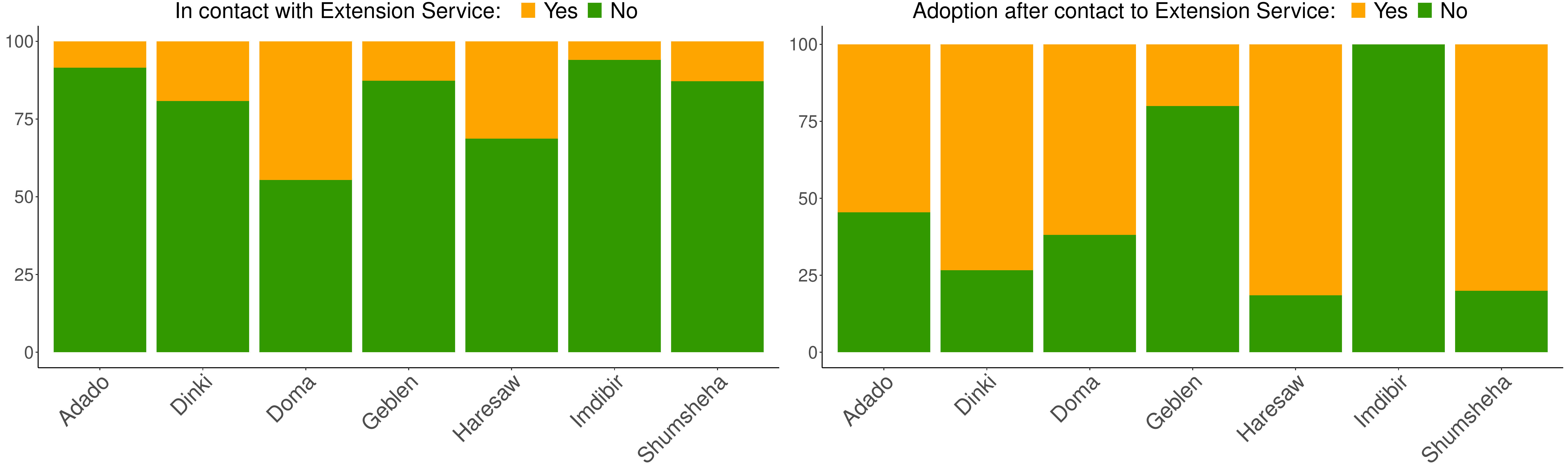}
\textit{Source:} Author's calculations based on ERHS. \\
\textit{Note:} Share of households by village that have been in contact with the extension service between 1994 and 2004 (left) and share of fertiliser adoptions in same time period conditional on the contact to extension agents (right).\\
\end{footnotesize}
\end{figure}

For the purpose of counteracting soil degeneration and promoting food stability in rural societies, extension service programmes aim to introduce modern inputs such as fertiliser and new agricultural practices. Extension service agents are present in each of the late blooming PAs. The data allows us to track extension service participation between 1994 and 2004 for the SG2000 and PADETES programmes. During that period, 166 out of 643 households adopted fertiliser and almost 50\,\% of adopters had had previous contact with the extension service program. Considering that only 18\,\% of households have had contact with the extension service, 71.55\,\% of these farmers started using fertiliser in the same year. \hyperref[Participation]{Figure\,\ref{Participation}} presents the extension service correspondence and subsequent adoption. While correspondence varies, impact appears to be quite positive for the majority of villages. Only households in Imdibir do not respond to the incentives given by extension agents. \par

\begin{figure}[H]
\begin{footnotesize}
 \caption[Most important activities of Extension Workers between 1994\,-\,1999]{\centering{\textit{Most important activities of Extension Workers between 1994\,-\,1999}}} 
 \label{impact_Ext} 
\hspace{2ex} 
\includegraphics[scale=0.4]{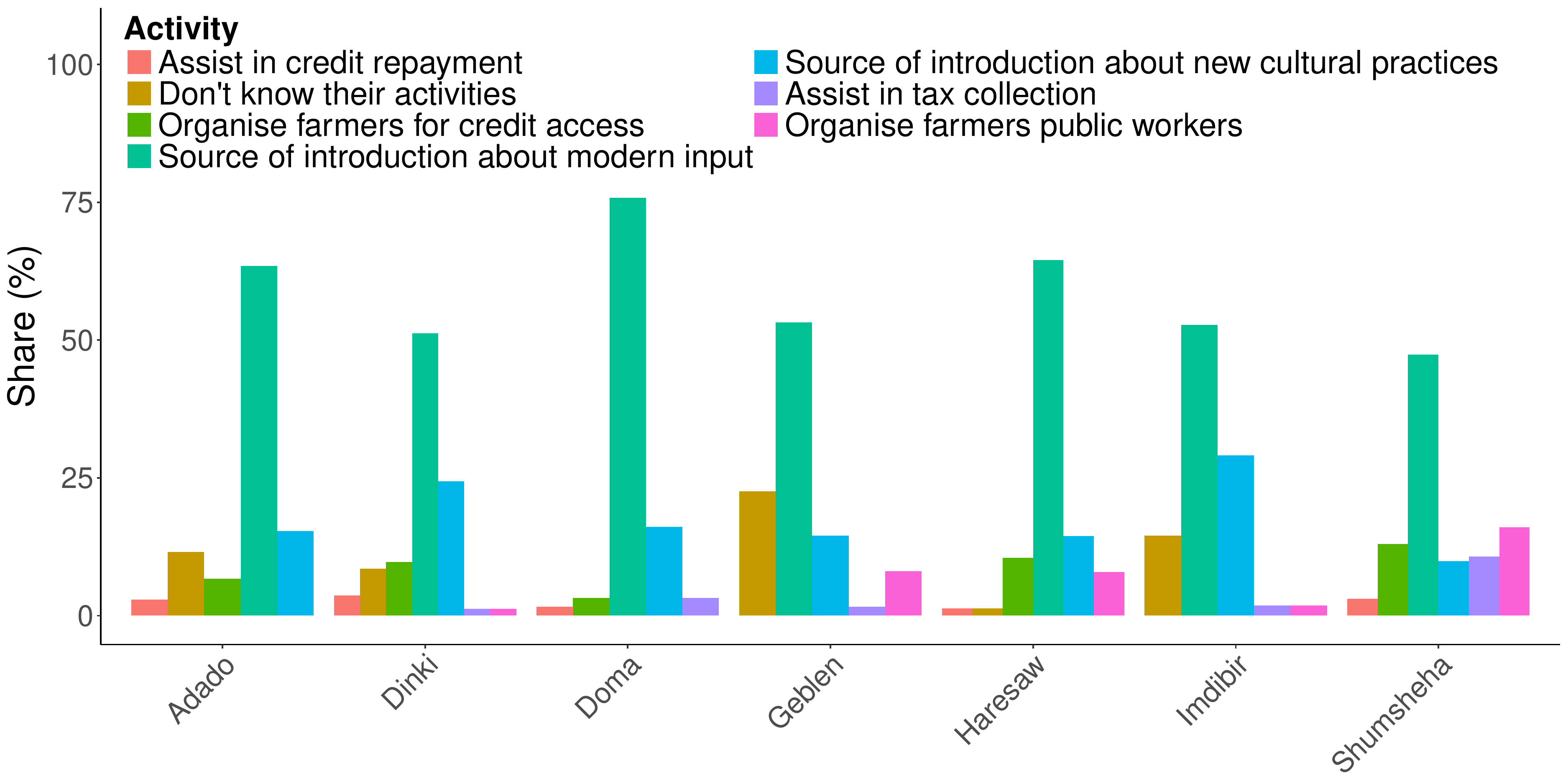} \\
\textit{Source:} Author's calculations, based on data from the ERHS.  
\end{footnotesize}
\end{figure}

Analysing supplementary data from the ERHS does not reveal systematic differences between PAs in terms of their understanding and acknowledgement of extension work. In all villages, the vast majority perceives the extension service as source of modern inputs and new cultural practices. Nonetheless, \hyperref[impact_Ext]{Appendix Figure\,\ref{impact_Ext}} also shows a reasonable share of missing awareness about the activities of extension agents for peasants in Adado, Geblen and Imdibir. The lack of awareness might explain the low participation shares seen in \hyperref[Participation]{Figure\,\ref{Participation}}. \par

Further reasons to not participate may originate from the design of the programmes which require a down payment and sufficient land size. Most farmers indicate one of both as the main barrier to participate (see \hyperref[non_Ext]{Appendix Figure\,\ref{non_Ext}}). Surprisingly, peasants in Imdibir neither perceive their small plot sizes nor the down payment as a main obstacle to joining the program. Instead they do not specify the main reason. \par 

In what follows, we will consider these aspects to properly estimate the impact of the extension service on fertiliser adoption among the selected villages.

\section{Methodology}
\label{Method}
\subsection{Estimation strategy}

Participation in an extension service programme or the seeking of advice from the extension agent suffers from self-selection bias, if more feasible farmers in terms of wealth, education and skills have higher odds of getting into contact with the development agent \citep{Dercon2009}. The probability of selection-bias is relatively high for the present data as extension agents require the achievement of a minimum adoption quota and therefore may target farmers with, a priori, superior conditions for adopting \citep{Belay2003}. Thus, estimating the impact of the extension service on fertiliser adoption demands adjustment as treated and non-treated farmers are assumed to behave differently even in the absence of the treatment. In an optimal scenario, we would like to measure the average treatment effect on the treated group: 

 \begin{equation} 
ATT = E\,[\,Y^T - Y^C \mid T\,]
\end{equation} 

\noindent with $Y^T$ as adoption of fertiliser (outcome variable) if treated and $Y^C$ if not treated, under the condition that all farmers belong to treatment group T. Clearly, it is impossible to observe $E\,[\,Y^C\,\mid T\,]$. The problem is known as the unobservable counterfactual \citep{Holland1986}, i.e. how would the treated farmer behave in the absence of the treatment. Common approaches to encounter the issue are matching techniques. The basic idea is to create an artificial counterfactual by selecting farmers from the observed non-treated group which share the largest common overlap with treated farmers with respect to known determinants of the outcome variable and the assignment to treatment. Ideally, the artificial counterfactual group is (almost) identical in observable variables to the treated farmers. This case would eliminate the selection bias as treated and control units coincide in their pre-treatment attributes and allows the estimation of the ATT defined as follows:

\begin{equation} 
ATT = E\,[\,Y^T \mid T,\,X\,] - E\,[\,Y^C \mid C,\,X\,]
\end{equation} 

\noindent where X is the vector of covariates known to influence the outcome variable or the assignment to treatment.

\subsection{Matching Frontier}

The common technique for estimating ATT by using a matching procedure is propensity score matching which is widely applied across various fields of research. However, propensity score matching, like other matching techniques, suffers from the trade-off to optimize either the imbalance between treated and control group to reduce model dependence and bias on the one hand or keeping a sufficiently large sample size to restrain the variance of the estimate on the other hand. So far, joint optimization of sample size and balance has occurred manually as present matching techniques could only optimize one or the other. The recent approach of matching frontier by \citet{King2016} disentangles the trade-off between imbalance and variance by automatic and simultaneous optimization of sample size and balance \citep{King2016}. \par

To achieve the simultaneous optimization, matching frontier uses a greedy algorithm to encounter the high computational complexity. Their greedy algorithm is an iterative process that stepwise prunes observations with the maximum distance from the overlap area while considering an updated distance calculation at each iteration. The outcome is the matching frontier presenting all optimal matched sample subsets along the imbalance-variance trade-off, i.e. each matched subset provides the minimal imbalance given the number of pruned observations\footnote{\citet{King2016} proof the optimality of their algorithm as pruning does not affect the nearest matched unit(s). See \citet{King2016} for the full proof.}. The precise algorithm for achieving the matching frontier depends, inter alia, on the choice of the imbalance metric. We will apply two different imbalance metrics in combination with two different sets for pruning  to investigate the consistency of the ATT.  \par

The first parameter set to construct the matching frontier uses the discrete $L_{1}$ imbalance metric. The $L_{1}$ processes differences between treated and control units as bins of a multivariate histogram \citep{King2016}. The $L_{1}$ is defined as follows: 	

\begin{equation} 
L_{1}(H) = \frac{1}{2}  \sum_{(v_{1}...v_{k}) \in H}  \left| t_{v_{1}...v_{k}} - c_{v_{1}...v_{k}} \right|
\end{equation}

\noindent with $v_{1}$...$v_{k}$ as the range of covariates of the multivariate histogram and with the relative proportions\footnote{The proportions refer to the frequency of treated (control) units per stratum in relation to the entire frequency of treated (control) units in the strata \citep{King2016}.} of treated ($t_{v_{1}...v_{k}}$) and control ($c_{v_{1}...v_{k}}$) units by covariate. In combination with the $L_{1}$ metric no treated units will be pruned. Thus, the reduction in imbalance may appear more challenging if treated units turn out to be very distinct from the controls and finding suitable matches could become cumbersome. The optimal algorithm will iteratively prune observations with the largest proportional distance between treated and control units unless the imbalance metric is larger than in the previous repetition \citep{King2016}. \par

Contrary to the first set of parameters, the Average Mahalanobis Imbalance (AMI) metric and pruning of treated observations defines the second specification. The AMI measures the average of shortest distances between all units $i$ of a sample and their nearest unit in the opposite group by 

\begin{equation}
D = mean_{i}\left[D(X_{i},X_{j(i)})\right]
\end{equation}

\noindent with $X_{j(i)}$ as defined as

\begin{equation}
X_{j(i)} = \arg \min_{X_{j}\mid j \in \left\{1-T_{i}\right\}} \left[\sqrt{(X_{i} - X_{j})S^{-1}(X_{i} - X_{j})}\right]
\end{equation}

\noindent $\sqrt{(X_{i} - X_{j})S^{-1}(X_{i} - X_{j})}$ presents the distance between the multi-dimensional vectors $X_{i}$ and $X_{j}$ with $S$ as covariance matrix \citep{King2016}. 

Since the second specification allows pruning of unfeasible treated units, the estimation will not present the ATT of the sample but the treatment effect on feasible observations and it is crucial to interpret the estimate under consideration of matched sample characteristics, i.e. one should take into account which kind of observations have been pruned and which remain. 
The greedy matching algorithm would, starting from the full sample, match each control observation to the nearest treatment observation and provide the shortest distance for each match. Subsequently, the AMI calculates the average distance of all matches and units with the largest minimum distance to their matched counterpart are removed. The outcome is a reduced subset of the original data. If the reduced subset contains more than two observations and an AMI greater than zero, the procedure repeats but starting from the reduced subset until only two observations remain or the AMI becomes zero. The greedy algorithm produces optimal results since re-matching would not decrease distance \citep{King2016}. \par 

After defining the estimation strategy and the matching frontier to estimate the ATT, the following subsection will briefly describe the construction of the sample and provides insights into the performance of the chosen matching techniques on the sample selection. 

\subsection{Sample selection}

In order to estimate the impact of the extension service on fertiliser adoption, the sample is split into treated and control groups according to  household participation in the extension service programme. Since the treatment status can be observed over the time period spanning 1994 to 2004, the sample splits treatment and control groups for each year. The control group contains all households that never had contact with an extension agent throughout this period. These households appear in the control group each year as long they do not migrate or adopt fertiliser\footnote{Fertiliser adopters do not appear in the sample after the adoption took place as the concern of the work is to evaluate determinants of the very first adoption and not to address the repeated confirmation of the adoption decision.}. Additionally, households that receive treatment join the control group in all years prior to the year of treatment, i.e. households with treatment in 1997 can be part of the control group in 1994,\,1995,\,1996. The treatment group comprises only households that obtain an extension service in the same year. Once a household has been treated, it cannot serve as control in successive rounds. \par
Moreover, to avoid spurious results, treated units are removed from the sample in the years following their first treatment. In that way, households receiving repeated treatment over time or adopting at a later date are just present in the year(s) before obtaining treatment and in the year of treatment. The reasoning behind this is to avoid measurement error as households being treated are systematically different from the control units due to the treatment and their decision to adopt at a later stage may still be influenced by the treatment received in a prior round. Equally, households receiving multiple treatments cannot be matched properly after their first treatment as the repeated assignment to the treatment group is not independent from the previous event in the sample. Hence, the impact estimation is limited to measuring the immediate response of households to the extension service, i.e. decision about fertiliser adoption of a household within the same year of receiving the treatment. \par 
   
The matching occurs taking into consideration variables that are known - in the literature - to affect the adoption and/or the selection into the treatment group as described by \citet{Feder1985,Sunding2001,Croppenstedt2003, Asfaw2004, Dadi2004, Weir2004, Carlsson2005, Knowler2007, Duflo2011, Dercon2011, Krishnan2014}, in order to achieve a balanced sample with respect to the baseline characteristics of treatment and control group. Since matching requires complete information, pruning observations with missing values from original data reduce the number of observations to 643 unique households whereof 109 did receive treatment at a certain point in time between 1994 and 2004. With these households two samples are built to exploit the available observations and to circumvent information gaps of time variant variables during non survey rounds.\par

The first sample is a full pooling sample of 4728 unmatched observations and covers each year from 1994 to 2004. The full pooling sample contains non-survey years in order to include all 109 treated households. However, the inclusion of non-survey years in the full pooling sample restricts matching on time invariant variables\footnote{Farm size, literacy and sex are relatively time consistent but we account for potential changes. If it was impossible to determine the status of a variable with certainty for the point in time, the observation was excluded from the sample.}. In order to exploit also time variant variables a second sample of 1907 observations is built from the  1994, 1997, 1999 and 2004 survey rounds. This partial pooling sample allows to better match on potentially decisive factors but cuts half the number of observed treated units from 109 to 54. \par

\hyperref[Balance]{Table\,\ref{Balance}} presents the balance tests for the full pooling sample and the partial pooling sample. Balance is achieved for matched subsets of 282 and 104 observations with the $L_{1}$ and AMI in the full pooling sample. The partial pooling sample presents balance for subsets of 108 observations with both matching algorithms. A closer look on the unmatched full pooling samples reveals significant differences between treated and control units for household specific characteristics. Treated households have larger farm sizes, are mainly male headed and have higher levels of literacy. Surprisingly, households experiencing a worse environmental shock are more present in the treatment group. This observation is in line with the analysis of \citet{Marmai2016}, showing the influence of extreme weather shocks on innovative behaviour. However, both, the $L_{1}$ and AMI, reduce the imbalance between treated and control units to non-significant levels for the sub-samples drawn from the full data\footnote{The $L_{1}$ suffers from keeping all treated households and is not able to completely remove the significant differences in farm size.}. The partial pooling contains a larger set of variables but limits the sample to the 1994, 1997, 1999 and 2004 survey rounds. The additional variables are missing in the full pooling sample due to their unobservable specification during non-survey rounds. Besides being male and facing an environmental shock, treated units are younger and own an oxen which is also a crucial determinant of fertiliser adoption in Ethiopia \citep{Dadi2004}. Moreover, we do not observe significant differences in access to treatment between members belonging to the local religious or ethnic majority. The reason for including the membership indicator is twofold. First, \citet{Bekele2003} describes the insecurity in land tenure and resulting shorter planning horizons of farmers belonging, locally, to the ethnic minority in Ethiopia. They expect a positive correlation between the application of new conservation methods and the membership in the ethnic majority group. Against their hypothesis, members of the ethnic majority do have a positive and significant correlation with the non application of conservation techniques. The results of \citet{Bekele2003} show that different ethnic groups in the same area respond differently to new agricultural practices. Second, the characteristics of extension agents are unobserved and we cannot account if agents cultural background fit the local ethnic and religious environment. However, we assume agents to belong to the ethnic majority of the area as the distribution of governmental extension agents outside their home area could be counterproductive due to language barriers. Hence, matching on the majority status should account for cultural differences between extension agents and farmers. As hypothesized by \citet{Bekele2003}, the membership status should also account for land security and access to information, which besides farm size, includes off farm income and equb membership\footnote{Equb or iqub, are local types of rotating savings and credit associations.}, are important determinants of adoption in Ethiopia \citep{Bewket2007, Kassie2009, Abebaw2013}. \par

Matching occurs also for village characteristics in order to check for potential differences in supply and demand constraints. Distance to market measures the remoteness of a village and hence the dependence on the extension agent to access fertiliser. The fractionalization index for ethnicity and religion expresses the probability that two randomly drawn individuals do not have the same religion or ethnicity \citep{Alesina2003,Alesina2005}. It is a measure for the diversity of the rural society and is taken into account as fragmented societies are under suspicion of thwarting economic development due to lower provision of public goods, conflict, coordination troubles and ethnic favourism \citep{Horowitz1985,Easterly1997, LaPorta1999,Alesina2000, Alesina2003,Franck2012}. However, the combination of distance to market and both fractionalization indices makes it possible to correctly identify the village and therefore to consider the socio-environmental conditions for the matching process.   \par

\newgeometry{
  left=5mm,
  right=5mm,
  top=5mm,
  bottom=5mm,
}

\begin{landscape}
\thispagestyle{empty}

\begin{table}[htbp] 
\begin{footnotesize}
	\caption[Balancing tests for matched samples]{\centering{\textit{Balancing tests for matched samples}}}
 \label{Balance}  
\begin{tabularx}{\columnwidth}{@{} ll*{18}{Y} @{}}
	
\toprule
& Variable	          & \multicolumn{9}{c}{full pooling n = 4728} & \multicolumn{9}{c}{partial pooling n = 1907}  \\
                       
											\cmidrule(lr){3-11} \cmidrule(lr){12-20} 

&												&	 \multicolumn{3}{c}{Unmatched sample} 
                        &	 \multicolumn{3}{c}{$L_{1}$} 
											  &	 \multicolumn{3}{c}{AMI} 
												&	 \multicolumn{3}{c}{Unmatched sample}
                        &	 \multicolumn{3}{c}{$L_{1}$}
											  &	 \multicolumn{3}{c}{AMI} \\
											
	                     \cmidrule(lr){3-5}
											 \cmidrule(lr){6-8}  
											 \cmidrule(lr){9-11} 											
											 \cmidrule(lr){12-14}
											 \cmidrule(lr){15-17}  
											 \cmidrule(lr){18-20}

&             & C   & T  & Diff \textit{p.value}    & C & T & Diff \textit{p.value}			  & C & T & Diff \textit{p.value}
              & C   & T  & Diff \textit{p.value}    & C & T & Diff \textit{p.value}			  & C & T & Diff \textit{p.value} 
							\\						
\midrule 
\parbox[t]{2mm}{\multirow{26}{*}{\rotatebox[origin=c]{90}{Household characteristics}}} 	 

& Farm Size      
& 0.99    
& 1.27    
& \textbf{0.0291} 
& 1.02  
& 1.27  
& \textbf{0.0919} 
& 0.71  
& 0.91  
& 0.2974 

& 0.99    
& 1.25    
& 0.1827 
& 1.05  
& 1.25  
& 0.4147 
& 0.98  
& 0.93  
& 0.8235 
\\
& Sex   
&0.73   
&0.87    
& \textbf{0.0000} 
& 0.87  
& 0.87  
& 0.9753 
& 0.77  
& 0.79  
& 0.8440 

& 0.73    
& 0.85   
& \textbf{0.0178}
& 0.85  
& 0.85  
& 1.0000 
&0.84  
& 0.82  
& 0.8502 
\\
& Literacy 
&0.30   
&0.43    
& \textbf{0.0110} 
& 0.44  
& 0.43  
& 0.7960 
& 0.29  
& 0.34  
& 0.5743 

& 0.31    
& 0.41   
& 0.1359
& 0.52  
& 0.41  
& 0.2465 
&0.51  
& 0.46  
& 0.6667 
\\
& Shock   
&0.21   
&0.31    
& \textbf{0.0260} 
& 0.31  
& 0.31  
& 0.9220 
& 0.30  
& 0.34  
& 0.6870 

& 0.23     
& 0.37    
& \textbf{0.0438}
& 0.37   
& 0.37    
& 1.0000 
&0.49  
& 0.46  
& 0.8356 
\\
& Soil Quality  
&1.78   
&1.77    
&0.9111 
&  1.69  
& 1.77  
& 0.3892 
& 1.99  
& 1.88  
& 0.5189 

& 1.79     
& 1.88    
& 0.4481
& 1.94   
& 1.88     
& 0.6869 
&1.89  
&2.09  
& 0.2195 
\\
      
& Member Majority: & & & & & & & & & & & & & & & & & & \\
\\
& \hspace{4ex}Ethnicity  
&0.91   
&0.89    
& 0.4000
& 0.91 
& 0.89  
& 0.6379 
& 0.98   
& 0.97  
& 0.7144 

& 0.91     
& 0.93    
& 0.6946
& 0.98  
& 0.93   
& 0.1736 
&0.99  
& 0.96  
& 0.5436 
\\
& \hspace{4ex}Religion     
&0.73   
&0.72    
& 0.7654
& 0.75 
& 0.72  
& 0.5114 
& 0.79   
& 0.79  
& 0.9849 

& 0.73     
& 0.81    
& 0.1202
& 0.78  
& 0.81   
& 0.6366 
& 0.90  
& 0.86  
& 0.5725 
\\

& Equb Member
& &&&&&&&&

& 0.15     
& 0.20    
& 0.3367
& 0.09  
& 0.20   
& 0.1063 
& 0.11  
& 0.14  
& 0.6921 
\\ \\

& Age
&&&&&&&&&

& 48.14    
& 43.39    
& \textbf{0.0053}
& 41.63  
& 43.39   
& 0.4356 
& 43.01 
& 42.64  
& 0.8844 
\\ \\

& Remittance         
&&& && && & & 
& 0.11     
& 0.06   
& 0.1031
& 0.06   
& 0.06   
& 1.0000 
& 0.04 
& 0.07  
& 0.5334 
\\ \\

& Off Farm Income         
& &&&&&&&&

& 0.37     
& 0.30   
& 0.2408
& 0.41  
& 0.30   
& 0.2305 
& 0.31
& 0.39   
& 0.4586 
\\ \\

& Oxen Owned         
&&&&&&&&&
& 0.23     
& 0.43   
& \textbf{0.0050}
& 0.30  
& 0.43   
& 0.1638 
& 0.24 
& 0.29   
& 0.6295 
\\ 
&  & & & & & & & & & & & & & & & & & & \\
\midrule
\parbox[t]{2mm}{\multirow{10}{*}{\rotatebox[origin=c]{90}{PA characteristics}}} 
& Diffusion   
&0.12   
&0.12    
& 0.8996
& 0.11 
& 0.12  
& 0.6439 
& 0.07   
& 0.08  
& 0.7433 

& 0.10     
& 0.09    
& 0.2435
& 0.09  
& 0.09   
& 0.7526 
& 0.04   
& 0.04   
& 0.6803 
\\
& Distance to market
&13.06  
& 12.08  
& 0.2011
&  12.98 
&  12.08  
& 0.3535 
&14.48   
& 14.63 
& 0.9247 

& 13.03     
& 12.76   
& 0.7914
& 12.63  
& 12.76  
& 0.9284 
& 13.91  
& 14.25  
& 0.8361 
\\
& Fractionalization: & & & & & & & & & & & & & & & & & & \\
\\
& \hspace{4ex}Ethnic
&0.12   
&0.15    
& 0.1925
& 0.13 
& 0.15  
& 0.5474 
& 0.12   
& 0.11  
& 0.6617 

& 0.12           
& 0.13          
& 0.9117
& 0.11  
& 0.13   
& 0.6390 
& 0.09  
& 0.10  
& 0.6934 
\\
& \hspace{4ex}Religion 
&0.37   
&0.37    
& 0.9728
& 0.35 
& 0.37  
& 0.4347 
& 0.29   
& 0.29  
& 0.9998 

& 0.37           
& 0.33          
& 0.2291
& 0.34  
& 0.33   
& 0.8195 
& 0.31  
& 0.28  
& 0.6086 
\\

\midrule 
						
\end{tabularx}
\textit{Source:} Author's calculations, based on ERHS and using the MatchingFrontier by \citet{MatchingFrontier}. \\ 
\textit{Note:}  Bold p-value indicates differences are significant at a 10\% level or lower. The balance achieved for the full pooling sample is given by matched subsets of 282 and 104 observations for the $L_{1}$ and AMI respectively. Balance for the partial pooling sample is given by subsets of 108 observations for both matching algorithms.  \\

\end{footnotesize}
\end{table}

\end{landscape}

\restoregeometry

\section{Results and Discussion}
\label{Result}

\subsection{Average impact of access to agricultural extension on fertiliser adoption}
\hyperref[ATT]{Table\,\ref{ATT}} presents the estimated average effects of access to agricultural extension agents on chemical fertiliser adoption. The results across both matching techniques and for both data samples show positive and significant impacts on fertiliser adoption decision of households without prior fertiliser usage. Farmers with contact to extension agents are roughly 60\% more likely to adopt fertiliser in the same year for the complete data sample. Considering the reduced data sample but checking for a broader set of variables still reveals highly significant results and a large average effect. The substantial size of the effect reveals the importance of extension agents as sources of information and access to fertiliser for first-time adopters. The focus on the initial adoption of a household makes it possible to identify the pure impact of extension service as potential adopters cannot rely on their own experience and depend mainly on the performance of the extension agent. The remarkable effect of the analysis indicates that the extension service fulfils its function as gatekeeper and successfully introduces fertiliser into a rather unaware and inexperienced environment. \par

By definition the ATT estimated by the $L_{1}$ imbalance measure reveals the impact of extension service on all treated units. In contrast, we allow pruning of infeasible treated observation for the AMI. The specification translate into the emission of 71 treated households from the full pooling sample (38 of 109 treated households remain) and 26 treated households from the partial pooling sample (28 of 54 treated households remain). The elimination of treated households during the matching process	makes it possible to remove the significant unbalance in farm size, which did not vanish if the algorithm is forced to keep all treated units. \par

Pruning treated households changes the average characteristics of the treatment group and the ATT estimates on the AMI matched samples do not hold for pruned treated households. Recalling the mean values after matching from \hyperref[Balance]{Table\,\ref{Balance}}, we can conclude for our two AMI matched samples, that the ATT draws on households with farm sizes below one hectare and large distances to market. The households almost entirely belong to the ethnic majority and are more shock affected than pruned observations of the treatment group. 

\begin{table}[H]
\begin{small}
\caption[Effect of the agricultural extension service on fertiliser adoption]{\centering{\textit{Effect of the agricultural extension service on fertiliser adoption}}}
 \label{ATT}  
\begin{tabularx}{\textwidth}{@{} l*{4}{Y} @{}}
	
\toprule
 Outcome	             &	 \multicolumn{4}{c}{Treatment variable: contact to extension service} 	\\
	
	                                \cmidrule(lr){2-5}
																									             &	 \multicolumn{2}{c}{full pooling} & \multicolumn{2}{c}{partial pooling}	\\					
																															\cmidrule(lr){2-3} \cmidrule(lr){4-5}																			
			  & 		   $L_{1}$     &    AMI    
				& 		   $L_{1}$     &    AMI				\\

	\midrule 

Fertiliser adoption &  
0.616$^{***}$       &        0.632$^{***}$       &     
0.593$^{***}$       &    0.523$^{***}$                \\

        &  (0.039)     &        (0.060) 
				&  (0.071)     &    (0.060)						
																\\

Observations pre-Match & 
\multicolumn{2}{c}{4728} &
\multicolumn{2}{c}{1907} \\
%
 
treated pre-Match &
\multicolumn{2}{c}{109}& 
\multicolumn{2}{c}{54} \\ 
%
%

Observations post-Match &
282   &  104        	& 
108   &  108        	\\  

treated post-Match &
109   &    38       &	 
54   &    28       	\\ 	

\midrule 
						
\end{tabularx}
\begin{footnotesize}
\textit{Source:} Author's calculations, based on data from the ERHS. Calculations have been performed by the R package \textit{MatchingFrontier} by \citet{MatchingFrontier}. \\ 
\textit{Note:} According to \citet{King2016}, our chosen specification of the AMI estimates the feasible sample average treatment effect on the treated (FSATT) as pruning of treated units is allowed and the selected specification of the $L_{1}$ algorithm without pruning of treated observations estimates the sample average treatment effect on the treated (SATT). To comply with literature we use the common expression, average treatment effect on the treated (ATT), for both cases. Significant results for $^{*}$p$<$0.1; $^{**}$p$<$0.05; $^{***}$p$<$0.01. \\
\end{footnotesize}
\end{small}
\end{table}
\vspace{-3ex}

\begin{figure}[H]
\begin{footnotesize}
\caption[Performance of imbalance reduction algorithms]{\centering{\textit{Performance of imbalance reduction algorithms}}}
\label{Distance}
\vspace{-4ex}
\begin{multicols}{2}
\hspace{-1ex}
\includegraphics[scale=0.28]{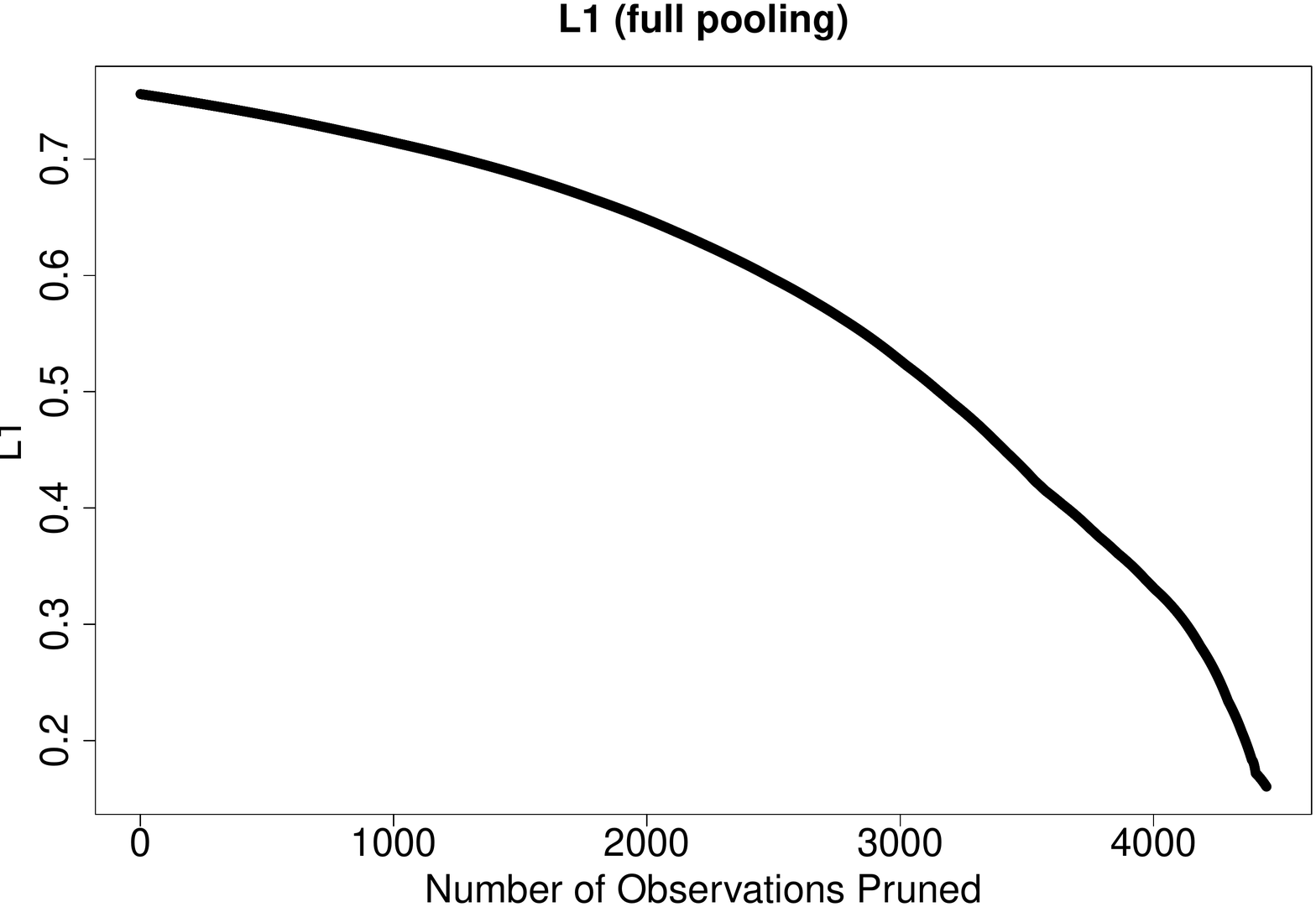}
\columnbreak
\includegraphics[scale=0.28]{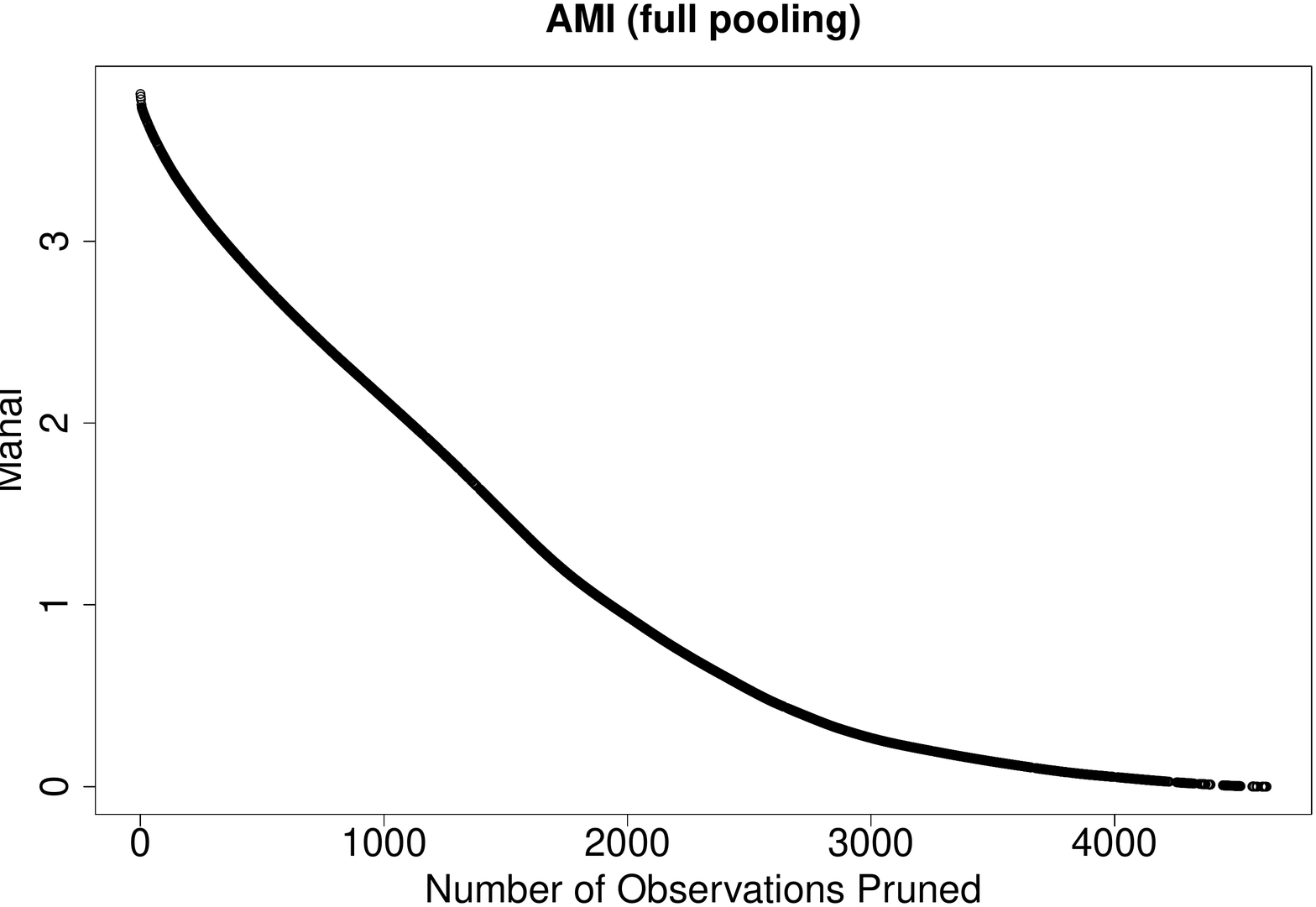}
\end{multicols}
\end{footnotesize}
\end{figure}

As shown in \hyperref[Balance]{Table\,\ref{Balance}}, significant differences between and treated and control units have been removed for both samples independent of the matching algorithm. \hyperref[Distance]{Figure\,\ref{Distance}} provides further information about the quality of pruning. The balanced sub-samples from the full pooling data sample portray the minimum distance achievable with both matching algorithms. In both cases the matching algorithm stopped as no further reduction in distance is possible, i.e. for full pooling: AMI = 0 and $L_{1}$ = 0.16. Given the characteristics of the matching frontier approach further pruning would not reduce imbalance. 

In contrast, both sub-samples from the partial pooling sample are optimal and balanced but do not present the lowest possible distance like the full pooling sub-samples due to the broader scope of variables and the lower number of observations. However, the sub-sample size of 108 observations is sufficient to remove significant differences between treated and control units. Also, the distance between treatment and control group decreased remarkably from an AMI of 11.404 with zero pruning to 0.404 for the selected sub-sample and from 0.97 to 0.445 for the $L_{1}$\footnote{Minimum distance achievable is zero for the AMI but requires pruning down to two observations and a $L_{1}$ distance of 0.428 for a sub-sample with more treated than control units.}. 

\vspace{-7ex}
\begin{figure}[H]
\begin{footnotesize}
\begin{multicols}{2}
\hspace{-1ex}
\includegraphics[scale=0.28]{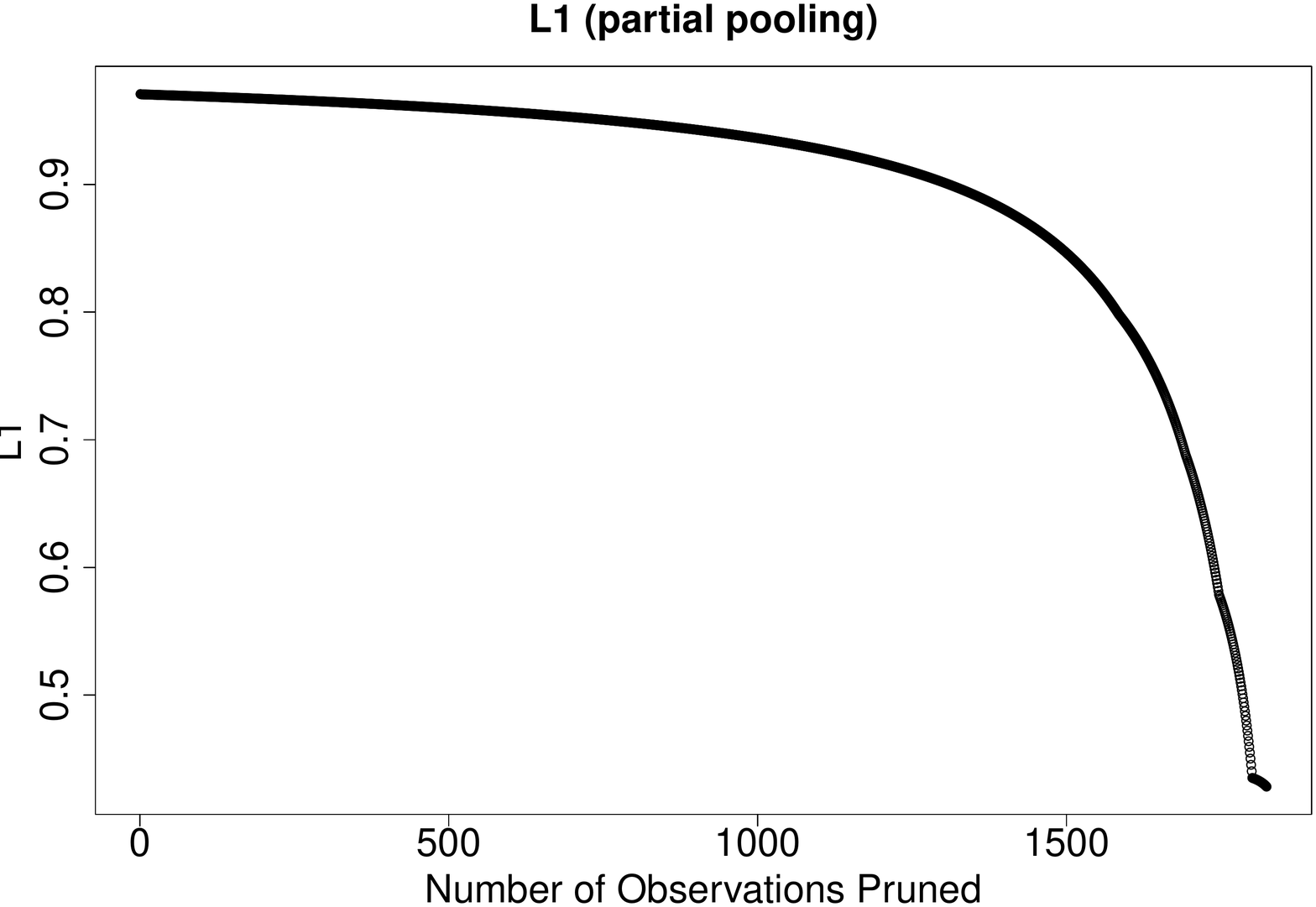}
\columnbreak
\includegraphics[scale=0.28]{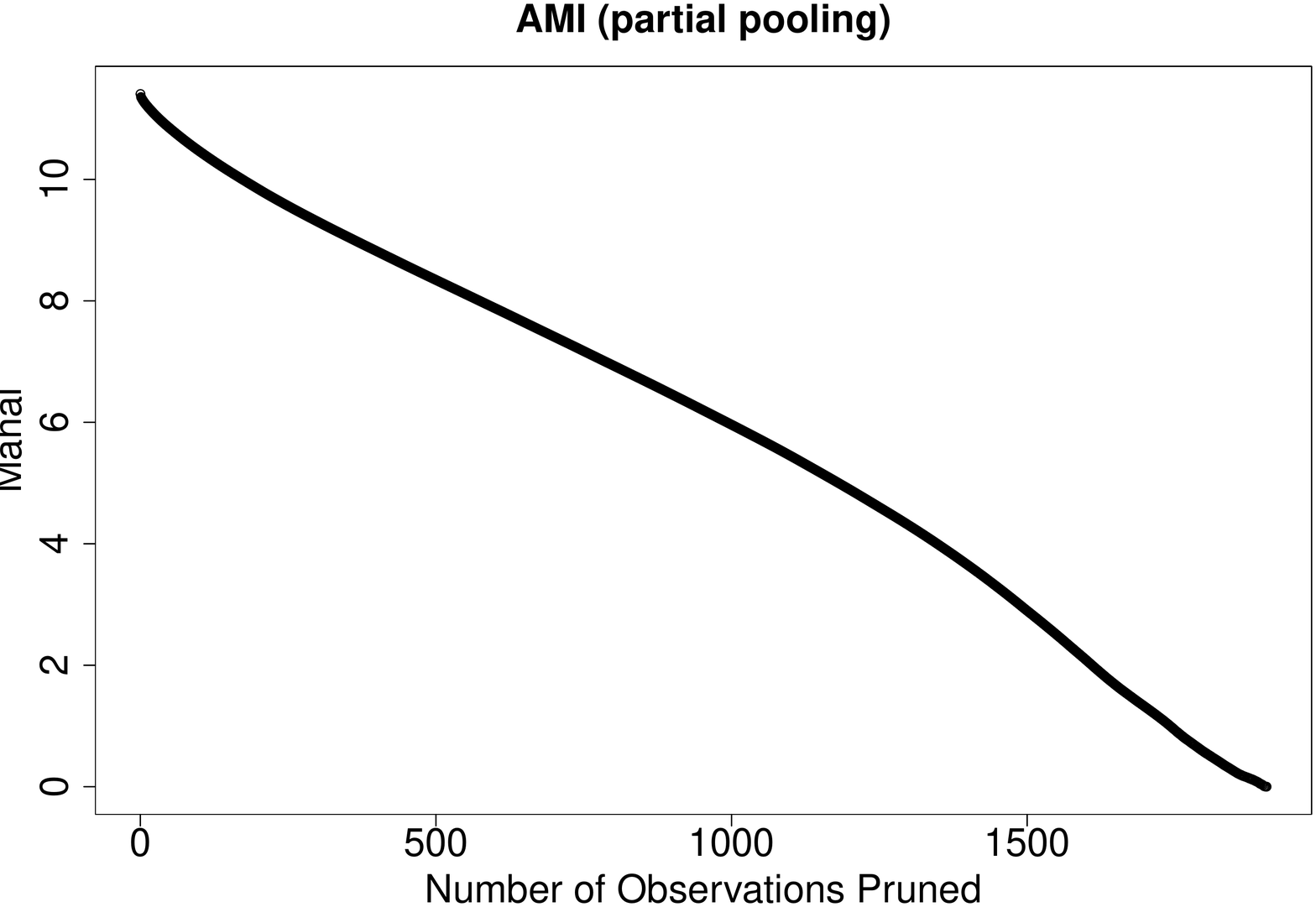}
\end{multicols}
\textit{Source:} Author's calculations, based on data from the ERHS. Calculations have been performed by the R package \textit{MatchingFrontier} by \citet{MatchingFrontier}. \\
\textit{Note:} Full pooling contains observations from every year between 1994 and 2004 but is limited to time invariant variables and variables with the possibility to calculate values for non-survey rounds, e.g. sex, farm size etc. Partial pooling uses observations from the ERHS 1994, 1997, 1999 and 2004 rounds and explores a broader set of variables.  \\
\end{footnotesize}
\end{figure}

\subsection{Robustness check}

In order to inspect the robustness of the estimated ATT, the matching procedure has been iterated for each year of the time period across all villages to analyse annual variations and to observe local differences relating to each village\footnote{All results depend on selected sub-samples that achieve a complete balance between treatment and control group.}. \par

Results for the annual variations, seen in \hyperref[YearMatch]{Table\,\ref{YearMatch}}, display the consistency of the effect over time. The largest impact coincides with the liberalisation and deregulation of fertiliser markets and prices in 1998 \citep{Abrar2004}. Notably the reported fade out of the effect in our sample begins as early as 2003\footnote{The effect was not measurable since no household received treatment in 2001 and not a single household adopted fertiliser for the first time in 2002.} and not only after 2004 as shown by \citet{Krishnan2014}. The lower impact in 2003 and the absence of the effect in 2004 do not stringently indicate the redundancy of the extension service due to increased awareness about fertiliser diffusion. While the extension service is mainly not present in villages with advanced fertiliser diffusion after 2000, 55.56\% of treatments occur in villages with low diffusion levels. Moreover, \hyperref[PAMatch]{Table\,\ref{PAMatch}} and \hyperref[PAMatch2]{Table\,\ref{PAMatch2}} reveal that the reduction of impact mostly reflects the poor extension performance in two villages in particular.\par   

Extension service properly works in Doma, Dinki, Haresaw and Shumsheha for full and partial pooling samples. The magnitude of the effect in these villages is partially larger than the main effect indicating a poorer performance in Adado, Geblen and Imdibir. Indeed, extension service fails to convince farmers to adopt fertiliser in the same year of the treatment in Geblen and Imdibir for each sample. In Adado, a positive and substantial effect is measurable for the full pooling sample. However, the impact vanishes for the reduced sample but cannot be attributed to the wider set of control variables in that sub-sample as accidentally no one of the three treated households in has adopted while five of six treated households have adopted in the years between the survey rounds. \par

The observed results have to be taken with caution. In particular, results from \hyperref[PAMatch2]{Table\,\ref{PAMatch2}} may suffer from the low number of remaining treated households. However, as the direction and magnitude of the effect does not differ in all but one village between \hyperref[PAMatch]{Table\,\ref{PAMatch}} and \hyperref[PAMatch2]{Table\,\ref{PAMatch2}}, we are confident about the impact of the extension service on fertiliser adoption in each village.  

\newpage
\newgeometry{
  left=20mm,
  right=5mm,
  top=7.5mm,
  bottom=7.5mm,
}
\begin{landscape}
\thispagestyle{empty}

\renewcommand{\arraystretch}{1.5}
\begin{table}[H]
\begin{footnotesize}
 \caption[Effect of the agricultural extension service on fertiliser adoption by year]{\centering{\textit{Effect of the agricultural extension service on fertiliser adoption by year}}} 
 \label{YearMatch}  
\begin{tabularx}{\columnwidth}{@{} l*{10}{Y} @{}}
	
\toprule
 Outcome	             &	 \multicolumn{10}{c}{Treatment variable: contact to extension service} 	\\
	
	                   \cmidrule(lr){2-11} 
																		
 &	
\multicolumn{2}{c}{1994} & 
\multicolumn{2}{c}{1995} & 
\multicolumn{2}{c}{1996} & 
\multicolumn{2}{c}{1997} & 
\multicolumn{2}{c}{1998} \\

\cmidrule(lr){2-3}		
\cmidrule(lr){4-5}		
\cmidrule(lr){6-7}		
\cmidrule(lr){8-9}		
\cmidrule(lr){10-11}		

&
$L_{1}$   &  AMI	& 	
$L_{1}$   &  AMI	& 	
$L_{1}$   &  AMI	& 	
$L_{1}$   &  AMI	& 	
$L_{1}$   &  AMI	\\

\midrule 

Fertiliser adoption &
0.545$^{***}$   &  0.471$^{***}$	& 	
0.404$^{***}$   &  0.397$^{***}$	& 	
0.667$^{***}$   &  0.800$^{***}$	& 	
0.755$^{***}$   &  0.636$^{***}$	& 	
0.989$^{***}$   &  0.956$^{***}$	\\ 	

 &
(0.094)   &  (0.087)	& 	
(0.097)   &  (0.080)	& 	
(0.054)   &  (0.046)	& 	
(0.062)   &  (0.099)	& 	
(0.028)   &  (0.058)	\\ 	

Observations pre-Match &
\multicolumn{2}{c}{630}& 	
\multicolumn{2}{c}{502}& 	
\multicolumn{2}{c}{483}& 	
\multicolumn{2}{c}{458}& 	
\multicolumn{2}{c}{389}\\ 	

treated pre-Match &
\multicolumn{2}{c}{22}& 	
\multicolumn{2}{c}{15}& 	
\multicolumn{2}{c}{6}& 	
\multicolumn{2}{c}{22}& 	
\multicolumn{2}{c}{14}\\ 	

Observations post-Match &
51   &  51        	& 	
63   &  64        	& 	
83   &  83        	& 	
80   &  79        	& 	
200   &  58        \\ 	

treated post-Match &
22   &    17       	& 	
15   &    12      	& 	
6  &      5    	& 	
22  &     17     	& 	
14   &    13      	\\ 	


 & & & & & & & & & &  \\
\midrule 
 Outcome	             &	 \multicolumn{10}{c}{Treatment variable: contact to extension service} 	\\
	
	                   \cmidrule(lr){2-11} 
& &
\multicolumn{2}{c}{1999}  &
\multicolumn{2}{c}{2000} & 
\multicolumn{2}{c}{2003} & 
\multicolumn{2}{c}{2004} \\ 
		
\cmidrule(lr){3-4}		
\cmidrule(lr){5-6}		
\cmidrule(lr){7-8}		
\cmidrule(lr){9-10}

& &
$L_{1}$   &  AMI	& 	
$L_{1}$   &  AMI	& 	
$L_{1}$   &  AMI	& 	
$L_{1}$   &  AMI	\\

\midrule

Fertiliser adoption & &
0.800$^{***}$   &  0.777$^{***}$	& 	
0.500$^{***}$   &  0.471$^{***}$	& 	
0.238$^{***}$   &  0.333$^{***}$	& 	
$-$0.006   &  $-$0.006	\\ 	

& &
(0.043)   &  (0.081)	& 	
(0.061)   &  (0.086)	& 	
(0.074)   &  (0.052)	& 	
(0.040)   &  (0.039)	\\ 	

Observations pre-Match & &
\multicolumn{2}{c}{399}& 	
\multicolumn{2}{c}{340}& 	
\multicolumn{2}{c}{315}& 	
\multicolumn{2}{c}{337}\\ 	

treated pre-Match & &
\multicolumn{2}{c}{5}& 	
\multicolumn{2}{c}{8}& 	
\multicolumn{2}{c}{4}& 	
\multicolumn{2}{c}{4}\\ 	

Observations post-Match & &
92   &  91        	& 	
77   &  77        	& 	
85   &  86        	& 	
159  & 171         	\\ 	

treated post-Match & &
5   &     5     	& 	
8  &    8      	& 	
4  &    3       & 	
4  &    4      	\\ 	

\midrule 
						
\end{tabularx}

\textit{Source:} Author's calculations, based on data from the ERHS. Calculations have been performed by the R package \textit{MatchingFrontier} by \citet{MatchingFrontier}. \\ 
\textit{Note:} Estimations not possible in the years 2001 and 2002 due to lack of treatment or adoption observations. The variables applied in each group have been presented in \hyperref[Balance]{Table\,\ref{Balance}}. The ERHS 1994, 1997, 1999 and 2004 rounds exploit the broader set of variables. Significant results for $^{*}$p$<$0.1; $^{**}$p$<$0.05; $^{***}$p$<$0.01 \\ 

\end{footnotesize}
\end{table}

\begin{table}[H]
\begin{footnotesize}
 \caption[Effect of the agricultural extension service on fertiliser adoption by village (full pooling)]{\centering{\textit{Effect of the agricultural extension service on fertiliser adoption by village (full pooling)}}} 
 \label{PAMatch}  
\begin{tabularx}{\columnwidth}{@{} l*{15}{Y} @{}}
	
\toprule
 Outcome	             &	 \multicolumn{14}{c}{Treatment variable: contact to extension service} 	\\
	
	                   \cmidrule(lr){2-15} 
																		
 &	
\multicolumn{2}{c}{Adado} & 
\multicolumn{2}{c}{Dinki} & 
\multicolumn{2}{c}{Doma} & 
\multicolumn{2}{c}{Geblen} & 
\multicolumn{2}{c}{Haresaw} & 
\multicolumn{2}{c}{Imdibir} & 
\multicolumn{2}{c}{Shumsheha} \\

\cmidrule(lr){2-3}		
\cmidrule(lr){4-5}		
\cmidrule(lr){6-7}		
\cmidrule(lr){8-9}		
\cmidrule(lr){10-11}		
\cmidrule(lr){12-13}		
\cmidrule(lr){14-15}		

&
$L_{1}$   &  AMI	& 	
$L_{1}$   &  AMI	& 	
$L_{1}$   &  AMI	& 	
$L_{1}$   &  AMI	& 	
$L_{1}$   &  AMI	& 	
$L_{1}$   &  AMI	& 	
$L_{1}$   &  AMI	\\

\midrule

Fertiliser adoption &
0.556$^{***}$   &  0.571$^{***}$	& 	
0.702$^{***}$   &  0.684$^{***}$	& 	
0.558$^{***}$   &  0.572$^{***}$	& 	
$-$0.060        &  $-$0.015 	& 	
0.800$^{***}$   &  0.827$^{***}$	& 	
$-$0.012        &  $-$0.025     	& 	
0.809$^{***}$   &  0.777$^{***}$	\\ 	

 &
(0.057)   &  (0.056)	& 	
(0.072)   &  (0.074)	& 	
(0.080)   &  (0.075)	& 	
(0.091)   &  (0.046)	& 	
(0.051)   &  (0.053)	& 	
(0.056)   &  (0.079)	& 	
(0.055)   &  (0.058)	\\ 	

Observations pre-Match &
\multicolumn{2}{c}{1086}& 	
\multicolumn{2}{c}{590}& 	
\multicolumn{2}{c}{388}& 	
\multicolumn{2}{c}{483}& 	
\multicolumn{2}{c}{544}& 	
\multicolumn{2}{c}{597}& 	
\multicolumn{2}{c}{1040}\\ 	

treated pre-Match &
\multicolumn{2}{c}{9}& 	
\multicolumn{2}{c}{15}& 	
\multicolumn{2}{c}{31}& 	
\multicolumn{2}{c}{7}& 	
\multicolumn{2}{c}{25}& 	
\multicolumn{2}{c}{4}& 	
\multicolumn{2}{c}{18}\\ 	

Observations post-Match &
87   &  87        	& 	
79   &  79        	& 	
75   &  75        	& 	
74   &  74        	& 	
87   &  87        	& 	
85   &  85        	& 	
101   &  101        	\\ 	

treated post-Match &
9   &  7        	& 	
15   &  14        	& 	
31   &  22        	& 	
7   &  7        	& 	
25   &  19        	& 	
4   &  4        	& 	
18   &  15        	\\ 	

\midrule 
						
\end{tabularx}

\textit{Source:} Author's calculations, based on data from the ERHS. Calculations have been performed by the R package \textit{MatchingFrontier} by \citet{MatchingFrontier}. \\ 
\textit{Note:} Significant results for $^{*}$p$<$0.1; $^{**}$p$<$0.05; $^{***}$p$<$0.01 \\ 

\end{footnotesize}
\end{table}


\begin{table}[H]
\begin{footnotesize}
 \caption[Effect of the agricultural extension service on fertiliser adoption by village (partial pooling)]{\centering{\textit{Effect of the agricultural extension service on fertiliser adoption by village (partial pooling)}}} 
 \label{PAMatch2}  
\begin{tabularx}{\columnwidth}{@{} l*{15}{Y} @{}}	
	
\toprule
 Outcome	             &	 \multicolumn{14}{c}{Treatment variable: contact to extension service} 	\\
	
	                   \cmidrule(lr){2-15} 
																		
 &	
\multicolumn{2}{c}{Adado} & 
\multicolumn{2}{c}{Dinki} & 
\multicolumn{2}{c}{Doma} & 
\multicolumn{2}{c}{Geblen} & 
\multicolumn{2}{c}{Haresaw} & 
\multicolumn{2}{c}{Imdibir} & 
\multicolumn{2}{c}{Shumsheha} \\

\cmidrule(lr){2-3}		
\cmidrule(lr){4-5}		
\cmidrule(lr){6-7}		
\cmidrule(lr){8-9}		
\cmidrule(lr){10-11}		
\cmidrule(lr){12-13}		
\cmidrule(lr){14-15}		

&
$L_{1}$   &  AMI	& 	
$L_{1}$   &  AMI	& 	
$L_{1}$   &  AMI	& 	
$L_{1}$   &  AMI	& 	
$L_{1}$   &  AMI	& 	
$L_{1}$   &  AMI	& 	
$L_{1}$   &  AMI	\\

\midrule

Fertiliser adoption &
$-$0.013        &  $-$0.017	& 	
0.667$^{***}$   &  0.729$^{***}$	& 	
0.750$^{***}$   &  0.727$^{***}$	& 	
$-$0.054        & $-$0.036 	      & 	
0.664$^{***}$   &  0.751$^{***}$	& 	
$-$0.014        &  $-$0.020     	& 	
0.979$^{***}$   &  0.959$^{***}$	\\ 	

 &
(0.067)   &  (0.076)	& 	
(0.071)   &  (0.097)	& 	
(0.067)   &  (0.068)	& 	
(0.115)   &  (0.094)	& 	
(0.093)   &  (0.085)	& 	
(0.069)   &  (0.082)	& 	
(0.056)   &  (0.090)	\\ 	

Observations pre-Match &
\multicolumn{2}{c}{429}& 	
\multicolumn{2}{c}{232}& 	
\multicolumn{2}{c}{166}& 	
\multicolumn{2}{c}{198}& 	
\multicolumn{2}{c}{220}& 	
\multicolumn{2}{c}{243}& 	
\multicolumn{2}{c}{419}\\ 	

treated pre-Match &
\multicolumn{2}{c}{3}& 	
\multicolumn{2}{c}{6}& 	
\multicolumn{2}{c}{12}& 	
\multicolumn{2}{c}{4}& 	
\multicolumn{2}{c}{19}& 	
\multicolumn{2}{c}{3}& 	
\multicolumn{2}{c}{7}\\ 	

Observations post-Match &
78  &  78        	& 	
52   &  52        	& 	
55  &  55        	& 	
60  &  60        	& 	
60   &  56        	& 	
73  &  53        	& 	
54   &  54        	\\ 	

treated post-Match &
3   &  3        	& 	
6   &  4        	& 	
12  &  11        	& 	
4   &  4        	& 	
19  &  15        	& 	
3   &  3        	& 	
7   &  5        	\\ 	

\midrule 
						
\end{tabularx}

\textit{Source:} Author's calculations, based on data from the ERHS. Calculations have been performed by the R package \textit{MatchingFrontier} by \citet{MatchingFrontier}. \\ 
\textit{Note:} Significant results for $^{*}$p$<$0.1; $^{**}$p$<$0.05; $^{***}$p$<$0.01 \\ 

\end{footnotesize}
\end{table}

\end{landscape}

\restoregeometry

\subsection{Impact heterogeneity}

The missing impact on fertiliser adoption in Geblen and Imdibir requires further investigation to understand potential barriers to the extension service system.  Notably, Geblen and Imdibir record the lowest number of treated households among all villages. Since we neither observe the supply of extension service, i.e. agents availability, nor the demand or willingness of farmers to get in contact with agents, we can only judge the selection of treated units. \par

In Geblen, out of our sample 10 households\footnote{Recall the construction of the data sample. Households that receive treatment in one year are not represented in subsequent years as we measure only the direct impact of extension service on the adoption decision. Hence, the sample contains only 10 out of 14 adoptions.} adopted fertiliser between 1994 and 2004 and seven households had contact with extension agents during the same period. Recalling \hyperref[Balance]{Table\,\ref{Balance}}, the major differences between treated and control units are observable for farm size, sex, literacy, shock, age and oxen ownership. In Geblen, predominantly men belonging to an ethnic and religious majority have contact with the extension agent (see \hyperref[Geb_Bal]{Table\,\ref{Geb_Bal}}). Treated farmers seem to be younger and have a higher level of literacy. Farm size, soil quality and the weather shock indicator do consistently differ on average between treated and control units. Interestingly, Tigrawain farmers (= ethnic majority) receive treatment but do not adopt while 60\% of the adopters belong to the Saho minority ethnic group. This observation may imply a cultural selection bias of the extension agent towards the Tigrawain people or the lack of interest by Saho to cooperate with the extension agent. Potential reasons may stem from the unobserved ethnicity of the extension agent. Since Geblen is located in the Tigray Region and Ethiopia is ruled by an Tigrawain government it is not impossible to imagine the appointment of mainly Tigrawain agents in that area and hence a potential favourism of the own ethnic group appears simply in order to avoid language barriers. \par 

\begin{table}[H]
\begin{footnotesize}
\caption[Pre-Balance between treated and control by year for Geblen]{\centering{\textit{Pre-Balance between treated and control by year for Geblen}}}
 \label{Geb_Bal}  
\begin{tabularx}{\textwidth}{@{} l*{8}{Y} @{}}
	
\toprule
 Variable	    &	 \multicolumn{8}{c}{Year of treatment} 	\\
	            \cmidrule(lr){2-9}
				&	 \multicolumn{2}{c}{1994} & \multicolumn{2}{c}{1995} & \multicolumn{2}{c}{1997} & \multicolumn{2}{c}{2004}	\\					
					\cmidrule(lr){2-3} \cmidrule(lr){4-5}				\cmidrule(lr){6-7} \cmidrule(lr){8-9}															
			  & 		   C     &    T    & 		   C     &    T  & 		   C     &    T  & 		   C     &    T				\\

	\midrule 
Fertiliser adoption &   0.00  &    0.00  &0.00    &   0.00 & \textbf{0.16} &     \textbf{0.00} & 0.00   &    0.00 \\
Farm Size      &         0.24 &     0.25 & \textbf{0.40}   &   \textbf{0.19} &0.34  &    0.56 & 0.38   &    0.50 \\
Sex            &         0.56 &     1.00 & 0.57   &   0.67 & \textbf{0.53}  &    \textbf{1.00} & 0.44  &     1.00 \\
Literacy       & 0.10 & 1.00	 & 0.09 & 0.33 & 0.08 & 0.50 & 0.15   &    1.00 \\
Shock          &         0.73  &    1.00 & 0.53 &   0.67 & 0.00   &   0.00 & 0.00 &      0.00 \\
Soil Quality   &         2.78  &    3.00 & 2.75 &   2.67 &\textbf{2.50}    &  \textbf{3.00} & 2.75  &     2.00 \\
Member Majority: &               &         & & & & &       &      \\
\hspace{4ex}Ethnicity &         0.66  &    1.00 & \textbf{0.64} &   \textbf{1.00} & \textbf{0.63}   &   \textbf{1.00}& 0.67   &    1.00 \\
\hspace{4ex}Religion &         0.79  &    1.00 & \textbf{0.79} &   \textbf{1.00} & \textbf{0.80}   &   \textbf{1.00}& 0.83   &    1.00 \\
Equb Member    &         0.00  &    0.00 &        & &   0.00  &    0.00 & 0.00  &     0.00 \\
Age            &         53.31 &    37.00&        & & \textbf{56.61}  &   \textbf{43.50} & 56.94 &     67.00 \\
Remittance     &         0.00  &    0.00 &        & &  0.02  &    0.00 & 0.64 &      1.00 \\
Off Farm Income&         0.81  &    1.00 &        & & \textbf{0.10}    &  \textbf{0.00} & 0.53    &   0.00 \\
Oxen Ownership &         0.02  &    0.00 &        & & \textbf{0.71}    &  \textbf{1.00} & 0.44   &    1.00 \\
Family Size    &         5.42  &    5.00 &        & & & & 4.86   &   10.00 \\
Trust          &               &         & & & & &        4.58  &     5.00  \\

\midrule 
						
\end{tabularx}
\textit{Source:} Author's calculations, based on data from the ERHS. Calculations have been performed by the R package \textit{MatchingFrontier} by \citet{MatchingFrontier}. \\ 
\textit{Note:} Bold values indicate differences are significant at a 10\% level or lower. Means in difference tests cannot be performed for 1994 and 2004 as each year contains only one treated observation. \\
\end{footnotesize}
\end{table}

To understand the direction of potential discrimination, further comparisons between treated households and Saho people, Tigrawai and Saho people as well as all non-treated households can be found in the \hyperref[SaTig]{Appendix Table\,\ref{SaTig}}. The comparisons reveal the following observations\footnote{The proposed explanations should be read with a great deal of caution as the comparisons suffer from the low number of treated observations.}. Saho people are by no means constantly worse than Tigrawai people in their pre-conditions for receiving treatment. Saho own slightly larger farms than Tigrawai farmers, have similar levels of literacy and soil quality, but household heads of Saho are somewhat older and families comprise on average almost one person more. Among all Tigrawai people the most feasible farmers received treatment, i.e. young farmers with relatively larger plots, higher literacy and oxen ownership. The comparison of treated farmers and Saho does not reveal a consistent superiority apart from the literacy. However, treated households are few in numbers and Saho farmers with identical attributes to treated Tigrawains could be selected. The selection towards treatment took place for farmers with suitable conditions to adopt fertiliser but unilateral selection of Tigrawains might not be by chance as the presence of ethnic blinkers cannot be determined with certainty. Inversely, generalized trust of Saho people is not significantly lower and cannot indicate a potential unwillingness to participate in the extension service. \par

Final comparisons of adopter and treated units shows partially larger farm sizes and a higher literacy by treated households (see \hyperref[Geb_Adopt]{Appendix Table\,\ref{Geb_Adopt}}). Adopters did not suffer from a weather shock in the year of adoption while treated households did face a weather shock at the same year. Since we cannot observe the exact point in time of the weather shock we do not know if the treatment occurs as a reaction to the shock. Interestingly, adopters of fertiliser differ partially in contrasting directions from their ethnic peers (see \hyperref[Geb_Eth]{Appendix Table\,\ref{Geb_Eth}}). While Saho adopters are the oldest farmers with the largest farms and the highest literacy, Tigrawain adopter are the youngest farmers with the smallest plots and the worst literacy. Both ethnic adopter groups neither earn off-farm income nor receive remittances but own oxen.  \par
\begin{figure}[H]
\begin{footnotesize}
 \caption[Reason to not participate Extension Service between 1994\,-\,1999]{\centering{\textit{Reason to not participate Extension Service between 1994 - 1999}}} 
 \label{non_Ext} 
\hspace{12ex} 
\includegraphics[scale=0.4]{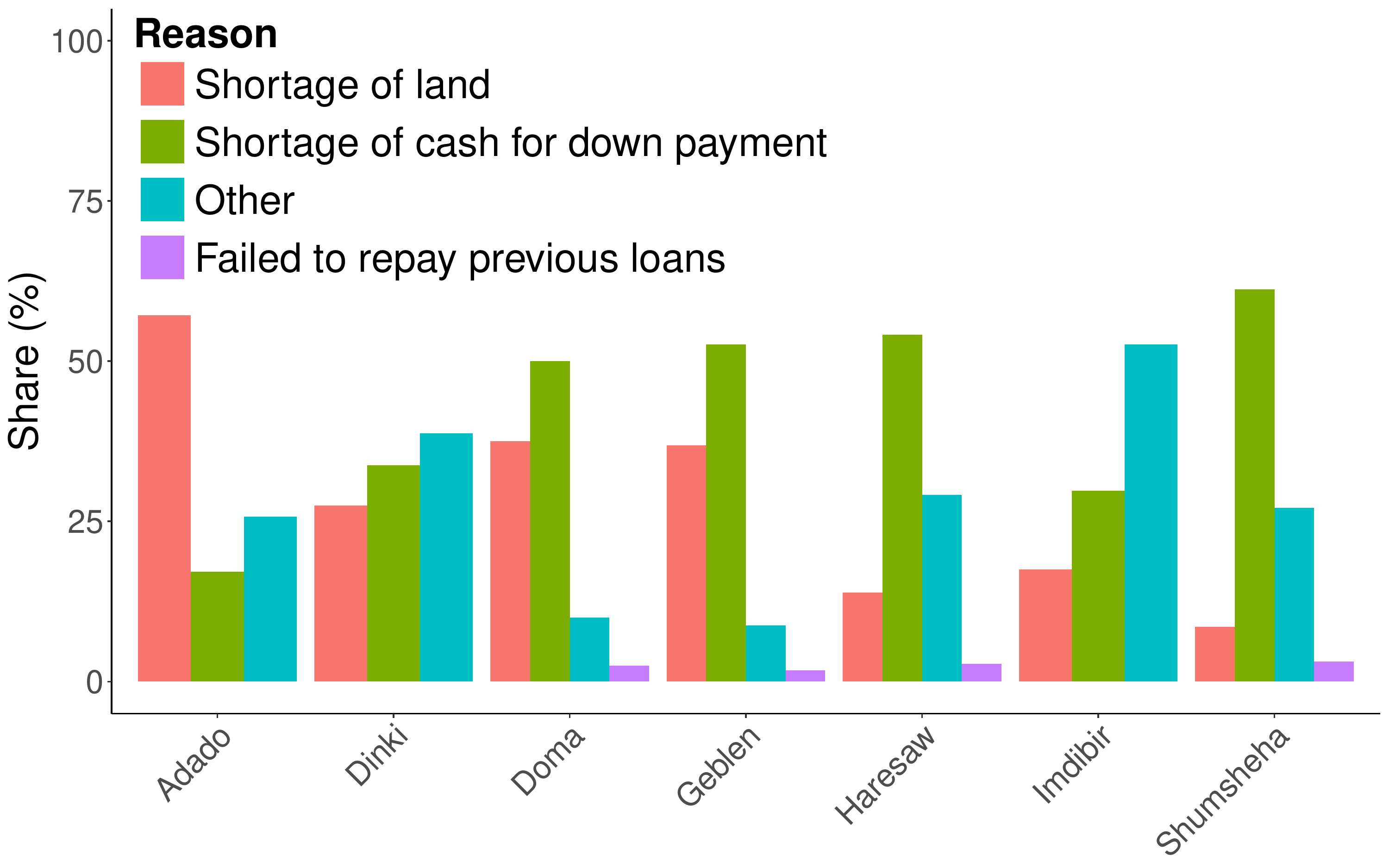} \\
\textit{Source:} Author's calculations, based on data from the ERHS. \\ 
\end{footnotesize}
\end{figure}

To sum up, the general selection mechanism in Geblen cannot be criticised based on the available data as treated units exhibit feasible economic pre-treatment characteristics to adopt fertiliser. Though a taste of discrimination either from Saho people towards the extension service or vice versa may stick. In addition, potential discrimination may also be present towards women as \citet{Ragasa2013} notice. In spite of the missing impact of extension service on adoption in the same year, sources for the general low participation should also briefly be mentioned. \hyperref[non_Ext]{Appendix Table\,\ref{non_Ext}} reveals the shortage of land and cash as main barriers to participate in programme between 1994 and 1999 in Geblen. However, the presented tables in the preceding paragraphs do not indicate superior pre-conditions for treated households in comparison to untreated with respect to farm size and income. A potential explanation for the low participation may be due to the limited availability of agents, as farmers in Geblen face the largest distance to market of all villages and agents may not show up regularly. \par

Imdibir, on the contrary, is only 5km away from the closest market. Yet, the village has the fewest adopter and extension service participants for the period being investigated. Moreover, the participation in extension service does not spur adoption. Pre-Balance comparisons between treated and controls in \hyperref[Imdibir]{Table\,\ref{Imdibir}} does not reveal a consistent superiority of economic characteristics of treated households. Only farm sizes of treated households are substantially larger in 2003 and 2004. Notably, all treated farmers are men. Comparing fertiliser adopter to treated households and in general to non-adopter in \hyperref[Imd_Adopt]{Table\,\ref{Imd_Adopt}} does not reveal significant differences for these groups. Since the economic variables available do not contribute to an understanding of the low participation and the missing impact a closer look certain distinctive socio-cultural attributes of Imdibir and its inhabitants may provide an answer. \par

\begin{table}[H]
\begin{footnotesize}
\caption[Pre-Balance between treated and control by year for Imdibir]{\centering{\textit{Pre-Balance between treated and control by year for Imdibir}}}
\label{Imdibir}  

\begin{tabularx}{\columnwidth}{@{} l*{6}{Y} @{}}
	
\toprule
 Variable	    &	 \multicolumn{6}{c}{Year of treatment} 	\\
	                  \cmidrule(lr){2-7}
				      &	 \multicolumn{2}{c}{1994} & \multicolumn{2}{c}{2003} & \multicolumn{2}{c}{2004}	\\					
					          \cmidrule(lr){2-3} \cmidrule(lr){4-5}				\cmidrule(lr){6-7}															
			               &  C    & T    &  C   &  T   &    C          &    T				\\

\midrule 
									
Fertiliser adoption  & 0.00  & 0.00 & 0.02 & 0.00 & 0.02          & 0.00 \\
Farm Size            & 0.15  & 0.09 & 0.24 & 0.44 & \textbf{0.24} & \textbf{0.38} \\
Sex                  & 0.86  & 1.00 & 0.70 & 1.00 & \textbf{0.69} & \textbf{1.00} \\
Literacy             & 0.37  & 1.00	& 0.35 & 0.00 & \textbf{0.36} & \textbf{0.00} \\
Shock                & 0.39  & 1.00 & 0.00 & 0.00 & 0.00          & 0.00 \\
Soil Quality         & 1.74  & 1.00 & 1.98 & 1.76 & 1.96          & 2.22 \\
Member Majority:     &       &      &      &      &               &      \\
\hspace{4ex}Religion & 0.48  & 0.00 & 0.43 & 0.00 & 0.53          & 0.50 \\
Equb Member          & 0.68  & 1.00 &      &      & 0.67          & 0.50 \\
Age                  & 49.23 & 35.00&      &      & 56.41         & 60.00 \\
Remittance           & 0.11  & 0.00 &      &      & \textbf{0.92} & \textbf{1.00} \\
Off Farm Income      & 0.61  & 1.00 &      &      & \textbf{0.47} & \textbf{0.00} \\
Oxen Ownership       & 0.00  & 0.00 &      &      & 0.00          & 0.00 \\
Family Size          & 7.59  & 8.00 &      &      & 5.35          & 6.50 \\
Trust                &       &      &      &      & 3.69          & 4.00  \\
\midrule 
						
\end{tabularx}
\textit{Source:} Author's calculations, based on data from the ERHS. Calculations have been performed by the R package \textit{MatchingFrontier} by \citet{MatchingFrontier}. \\ 
\textit{Note:} Bold values indicate differences are significant at a 10\% level or lower. Means in difference tests cannot be performed for 1994 and 2003 as each year contains only one treated observations. \\
\end{footnotesize}
\end{table}
      
Unlike in Geblen, economic barriers have not been stated as main reason for non-participation (see \hyperref[non_Ext]{Appendix Table\,\ref{non_Ext}}). Instead, almost 50\% of the peasants in Imdibir call the unspecified factor ``Other'' as a reason for rejecting participation. One interpretation of the factor ``Other'' could be the common low level of trust. In a comparison of different trust dimensions between all villages in \hyperref[Trust]{Appendix Table\,\ref{Trust}}, inhabitants of Imdibir present the poorest trust towards other in all categories. While the minor trust in people and neighbours is at least close to a neutral value, i.e. neither agree nor disagree to trust, the attitude towards the Ethiopian government and their local kebele government is illustrated by large mistrust and complicates the mission of the extension agents as governmental servants. \par

Also \citet{Imdibir} observe the suspiciousness of Gurage people (ethnic group of Imdibir) towards local outsiders. The Gurage community maintains close bonds and refuses to report any financial information due to fear of tax collection \citep{Imdibir}. The strong reliance on cultural traditions creates a common identity but can be an obstacle to foreigners with a different background to achieve acceptance by the community as firstly described by \citet{Wellin1955}. Hence, the work of extension agents may be hampered by a missing acceptance of the rural society and low willingness to collaborate due to the deficit in trust. Another barrier could stem from the broad dimension of religious beliefs in the Gurage society. This variety comes along with numerous holy days every month preventing farmers from work on their plots and limit potential visits of extension agents especially if the agents suspend work on other days due to religious mismatch. \par

A final note on the selection in both villages concerns the tendency to treat mainly younger farmers. \citet{Koutsou2014} show that younger farmers have lower levels of trust to people and are less innovative. Indeed, for the limited number of observations young farmers present lower levels of trust. Thus, fertiliser adoption may be hampered due to the missing trust of young farmers in the extension worker. The line of thoughts receives support by \citet{Abebe2016}, revealing that older farmers using middlemen for trading purposes are more frequent.

\section{Conclusion}
\label{ConclusionMatch}

The paper provides additional evidence for the importance of extension services programmes to the discussion about their efficiency. The performance of the extension service is estimated in the context of a late blooming environment, i.e. in contrast to other works, we consider 
only the first fertiliser usage as an adoption and focus on households in villages with low fertiliser diffusion during the time of the study launch in 1994. The restriction of the adoption definition implies that the household has never employed fertiliser before and hence cannot rely on previous personal experience during the adoption decision process. Our aim is to avoid spurious measurement of the extension service impact, as repeated fertiliser adoption of households over time potentially depends to a diminishing degree on extension service but may substituted by neighbours or own experiences.\par

Our results confirm the expected impact of the extension service on the first fertiliser adoption and prove their importance to introduce new ideas, raise awareness and foster adoption. The effect appears to be relatively constant over time but differs between villages. While the impact is large in magnitude for farmers in Dinki, Doma, Haresaw and Shumsheha, extension agents fail to immediately spur adoption in Geblen and Imdibir. \par

Attempts to explain the poor performance in both villages additionally raise the question about the generally low participation rates in these PA. Since we cannot observe the frequency and duration of extension agents visits in each village, inadequate supply provides a (non-testable) explanation. On the demand side, peasants in Imdibir reveal exceptionally low levels of trust in local and national authorities. The lack of trust may impede an appropriate collaboration and potentially explains the low participation levels as well as the missing impact on adoption. On the contrary, several farmers in Geblen adopt fertiliser but none of them due to extension service. The shortage of impact could not be explained by observable economic or social variables. \par

Hence, further research is required to evaluate the described channels as the proposed explanations suffer from missing data.

\newpage

\section*{Appendix}
\addcontentsline{toc}{section}{Appendix}

\vspace{2ex}

\vspace{-2ex}

\newpage
\newgeometry{
  left=20mm,
  right=5mm,
  top=7.5mm,
  bottom=7.5mm,
}
\begin{landscape}
\thispagestyle{empty}

\renewcommand{\arraystretch}{1.5}
\begin{table}[H]
\begin{footnotesize}
 \caption[Differences between Saho and Tigrawai people in Geblen]{\centering{\textit{Differences between Saho and Tigrawai people in Geblen}}} 
 \label{SaTig}  
\begin{tabularx}{\columnwidth}{@{} l*{16}{Y} @{}}
	
\toprule
 Variable	    &	 \multicolumn{16}{c}{Year of treatment} 	\\
	            \cmidrule(lr){2-17}
				&	 \multicolumn{4}{c}{1994} & \multicolumn{4}{c}{1995} & \multicolumn{4}{c}{1997} & \multicolumn{4}{c}{2004}	\\					
					         \cmidrule(lr){2-5} \cmidrule(lr){6-9}				\cmidrule(lr){10-13} \cmidrule(lr){14-17}

   			& 	Saho	&   \multicolumn{3}{c}{Tigrawai}     				 
				& 	Saho	&   \multicolumn{3}{c}{Tigrawai} 
				& 	Saho	&   \multicolumn{3}{c}{Tigrawai} 
				& 	Saho	&   \multicolumn{3}{c}{Tigrawai} 				\\ 

      \cmidrule(lr){3-5}  \cmidrule(lr){7-9}				\cmidrule(lr){11-13} \cmidrule(lr){15-17}
				& 		&   all     & treated     &    non-treated   				 
				& 		&   all     & treated      &    non-treated  
				& 		&   all     & treated      &    non-treated  
				& 		&   all     & treated      &    non-treated  				\\ 

\midrule 

Fertiliser adoption  &
0.00  &    0.00      &    0.00  & 0.00  &     
0.00  & 0.00          &  0.00     &  0.00 &     
0.26   &  0.09    &   \textbf{0.00}    &  0.09  &     
0.00   &  0.00         &  0.00     & 0.00  \\     

Farm Size    &      
0.30  & \textbf{0.22}  &  0.25  &  \textbf{0.22} & 
0.45    &    0.36       &  \textbf{0.19}     &  0.37 &     
0.34   & 0.35          &   \textbf{0.56}    & 0.34   &     
0.46   &  0.35         &  0.50     &  0.35 \\     
	
Sex &
0.71 &     0.50  &    1.00  &   \textbf{0.49} &  
0.68      &  0.51         &  0.67     & 0.50  &     
0.63   &   0.50        &  \textbf{1.00}     & 0.47   &     
0.50   &   0.44        &  1.00     &  0.42 \\     
	
Literacy &  
0.10  &    0.12  &  1.00   & 0.10 & 
0.11      &    0.11        &  0.33     & 0.09  &     
0.11   &   0.09        &  0.50     & 0.06   &     
0.17   &   0.18        &  1.00     & 0.15  \\     

Shock    &
0.62     &  0.79  & 1.00 & 0.78 & 
0.79      & \textbf{0.41}         &   0.67    & \textbf{0.38}  &     
0.00   & 0.00          &   0.00    & 0.00   &     
0.00   & 0.00          &  0.00     &  0.00  \\     
	
Soil Quality &
 2.86   &   2.75  & 3.00 &  2.74 & 
 2.74     &  2.75         &  2.67     & 2.75  &     
2.48   &    2.54       &  \textbf{3.00}     & 2.51   &     
2.76   &  2.71         &   2.00    &  2.74 \\     
	
Equb Member &        
0.00  &    0.00   & 0.00 & 0.00 & 
      &           &       &   &     
0.00   & 0.00      & 0.00      & 0.00   &     
 0.00  & 0.00          &  0.00     & 0.00  \\     
	
Age &
57.43  &   50.86  & 37.00 & 51.20 & 
      &           &       &   &     
 59.89  &  54.00     & \textbf{43.50}      & 54.66  &     
56.67   &  57.48         &  67.00     & 57.08  \\     
	
Remittance  &
0.00   &   0.00   &  0.00 & 0.00 & 
      &           &       &   &     
0.00   &   0.03     &  0.00     & 0.03   &     
0.67   &    0.64        &  1.00     &  0.62 \\     
	
Off Farm Income   &  
0.81  &    0.81   & 1.00 & 0.80 & 
      &           &       &   &     
0.00   &  \textbf{0.15}  & 0.00       & \textbf{0.16}   &     
 0.67   &   0.44        &  0.00     & 0.46  \\     
	
Oxen Ownership   &
0.05   &   0.00  &  0.00 & 0.00 & 
      &           &       &   &     
0.74   &  0.71      & \textbf{1.00}      &  0.69  &     
0.33   &   0.52        &  1.00     & 0.50  \\     
	
Family Size     &
5.86  &    5.19  & 5.00 & 5.20  & 
      &           &       &   &     
   &           &       &    &     
 5.50  &   4.76        &    10.00   &  4.54 \\     
	
Trust     &
       &      &  &   & 
      &           &       &   &     
   &           &       &    &     
4.50  &   4.64        & 5.00      &  4.62 \\     

\midrule 
		
\end{tabularx}
\textit{Source:} Author's calculations, based on data from the ERHS. \\ 
\textit{Note:} Bold values indicate  differences are significant at a 10\% level or lower in respect to the Saho value. Means in difference tests cannot be performed for column \textit{Tigrawai treated} in 1994 and 2004 as each year contains only one treated observation. \\
\end{footnotesize}
\end{table}

\end{landscape}
\restoregeometry

\begin{table}[H]
\begin{footnotesize}
\caption[Comparison of Adopter and Treated in Geblen]{\centering{\textit{Comparison of Adopter and Treated in Geblen}}}
 \label{Geb_Adopt}  
\begin{tabularx}{\columnwidth}{@{} l*{6}{Y} @{}}
	
\toprule
 Variable	 & \multicolumn{3}{c}{1997}   & \multicolumn{3}{c}{All years}  \\
            \cmidrule(lr){2-4}   \cmidrule(lr){5-7} 
   
	              &	 Adopter 	& Treated & Means difference test (\textit{p-value})   &	 Adopter 	& Treated & Means difference test (\textit{p-value})\\

	\midrule 
Farm Size       & 0.35  & 0.56  & \textbf{0.0927} & 0.33 & 0.35 & 0.8425\\
Sex             & 0.62  & 1.00  & \textbf{0.0796}          & 0.70 & 0.86 & 0.4643   \\
Literacy        & 0.12  & 0.50  & 0.5880          & 0.10 & 0.57 & \textbf{0.0662}\\
Shock           & 0.00  & 0.00  &                 & 0.00 & 0.43 & \textbf{0.0781}     \\
Soil Quality    & 2.00  & 3.00  & \textbf{0.0185}          & 2.20 & 2.71 & 0.1569     \\
Equb Member     & 0.00  & 0.00  &                 & &  & \\
Age             & 57.50 & 43.50 & \textbf{0.0221}          & &  & \\
Remittance      & 0.00  & 0.00  &                 & &  & \\
Off Farm Income & 0.12  & 0.00  & 0.3506          & &  & \\
Oxen Ownership  & 0.75  & 1.00  & 0.1705          & &  & \\

\midrule 
						
\end{tabularx}
\textit{Source:} Author's calculations, based on data from the ERHS. \\ 
\textit{Note:} Bold values indicate differences are significant at a 10\% level or lower. 1997 is the only year in which treated units and adopters are observed. However, no adopter received treatment and no treated household adopted in the same year.   \\
\end{footnotesize}
\end{table}

\begin{table}[H]
\begin{footnotesize}
\caption[Comparison of Adopter by Ethnicity for Geblen]{\centering{\textit{Comparison of Adopter by Ethnicity for Geblen}}}
 \label{Geb_Eth}  
\begin{tabularx}{\columnwidth}{@{} l*{6}{Y} @{}}
	
\toprule
 Variable	 & \multicolumn{3}{c}{Tigrawai}   & \multicolumn{3}{c}{Saho}  \\
            \cmidrule(lr){2-4}   \cmidrule(lr){5-7} 
   
	              &	 Adopter 	& Non-Adopter & Means difference test (\textit{p-value})   &	 Adopter 	& Non-Adopter & Means difference test (\textit{p-value})\\

	\midrule 
               
Farm Size       & 0.23  & 0.30  & 0.3059          & 0.40  & 0.33  & 0.2813\\
Sex             & 0.67  & 0.46  & 0.6011          & 0.67  & 0.61  & 0.7940   \\
Literacy        & 0.00  & 0.13  & \textbf{0.0000} & 0.17  & 0.11  & 0.7732\\
Shock           & 0.00  & 0.29  & \textbf{0.0000 }& 0.00  & 0.25  & \textbf{0.0000}     \\
Soil Quality    & 2.00  & 2.70  & 0.3515          & 2.17  & 2.77  & 0.1919     \\
Equb Member     & 0.00  & 0.00  &                 & 0.00  & 0.00  & \\
Age             & 53.33 & 54.49 & 0.8890          & 63.00 & 57.82 & 0.4568 \\
Remittance      & 0.00  & 0.14  & \textbf{0.0000} & 0.00  & 0.15  & \textbf{0.0021 }\\
Off Farm Income & 0.33  & 0.51  & 0.6540          & 0.00  & 0.52  & \textbf{0.0000} \\
Oxen Ownership  & 0.67  & 0.46  & 0.6011          & 0.83  & 0.41  & \textbf{0.0517} \\

\midrule 
						
\end{tabularx}
\textit{Source:} Author's calculations, based on data from the ERHS. \\ 
\textit{Note:} Bold values indicate differences are significant at a 10\% level or lower. Adoption mainly occurred in 1997 and 1999. Another adoption in 2003 has not been considered in the table to exploit a broader range of variables. \\
\end{footnotesize}
\end{table}

\begin{table}[H]
\begin{footnotesize}
\caption[Comparison of Adopter and Treated in Imdibir]{\centering{\textit{Comparison of Adopter and Treated in Imdibir}}}
 \label{Imd_Adopt}  
\begin{tabularx}{\columnwidth}{@{} l*{6}{Y} @{}}
	
\toprule
 Variable	 & \multicolumn{3}{c}{Treated vs. Adopter}   & \multicolumn{3}{c}{Adopter vs. Non-Adopter}  \\
            \cmidrule(lr){2-4}   \cmidrule(lr){5-7} 
   
	              &	 Treated 	& Adopter & Means difference test (\textit{p-value})   &	 Adopter 	& Non-Adopter & Means difference test (\textit{p-value})\\

	\midrule 
									
Farm Size       & 0.32  & 0.32  & 0.9527          & 0.32 & 0.25 & 0.4630\\
Sex             & 1.00  & 0.60  &  0.1778         & 0.60 & 0.78 & 0.5143   \\
Literacy        & 0.25  & 0.30  & 0.8809          & 0.30 & 0.37 & 0.7480 \\
Shock           & 0.25  & 0.20  & 0.8809          & 0.20 & 0.28 & 0.7263     \\
Soil Quality    & 1.80  & 1.82  & 0.9637          & 1.82 & 1.81 & 0.9807     \\

\midrule 
						
\end{tabularx}
\textit{Source:} Author's calculations, based on data from the ERHS. \\ 
\textit{Note:} Bold values indicate differences are significant at a 10\% level or lower. Adoption observed in 1995, 1997, 2003, 2004. Treatment observed in 1994, 2003, 2004.    \\
\end{footnotesize}
\end{table}

\begin{table}[H]
\begin{footnotesize}
\caption[Trust across Peasant Associations]{\centering{\textit{Trust across Peasant Associations}}}
 \label{Trust}  
\begin{tabularx}{\columnwidth}{@{} l*{7}{Y} @{}}
	
\toprule
 Trust in:	 & \multicolumn{7}{c}{Village}  \\
            \cmidrule(lr){2-8}    
   
		           &	 Adado 	& Dinki & Doma   &	 Geblen & Haresaw & Imdibir &  Shumsheha\\

		\midrule 
  People       &  4.79 &  5.15 &  4.75  &  4.64   &  5.04 &  3.58 &  4.45 \\
	Neighbours	&   5.48&    5.62 &    5.32&    5.00&    5.27&    4.25 &    5.18 \\
	Government	&    4.49	&    4.24&    5.18&    4.56&    4.97 &    2.55	&    4.72\\
	Kebele			&  4.14 &  4.04  &  4.57&  4.46 &  4.74 &  2.73 &  4.50 \\

\midrule 
						
\end{tabularx}
\textit{Source:} Author's calculations, based on data from the ERHS. \\ 
\textit{Note:} Trust measure base on a 7-point Likert scale. Higher values report stronger approval towards the trust statement. Value 1 corresponds to ``Strongly disagree'' and the value 7 corresponds to ``Strongly agree''. The Kebele represents the lowest level of local government.   \\
\end{footnotesize}
\end{table}

\newpage
\addcontentsline{toc}{section}{References}
\bibliographystyle{ametsoc2014}
\bibliography{Literature}

\end{document}